\documentclass[aps,prb,twocolumn,floatfix]{revtex4}
\usepackage{epsf}

\newcommand{\hybb}  {V_0}               
\newcommand{\hyb}   {v}                 
\newcommand{\epsfb} {\varepsilon_0}     
\newcommand{\epsf}  {\varepsilon}       
\newcommand{\Ub}    {U_0}               
\newcommand{\U}     {u}                 
\newcommand{\Jb}    {J_{\rm K}}               
\newcommand{\lam}   {\lambda_0}         
\newcommand{\SSS}   {\hat S}            
\newcommand{\TK}    {T_{\rm K}}         
\newcommand{\Db}    {\Delta_0}          

\begin{document}

\title{Impurity Quantum Phase Transitions} 

\author{Matthias Vojta}
\affiliation{Institut f\"ur Theorie der Kondensierten Materie,
Universit\"at Karlsruhe, Postfach 6980, 76128 Karlsruhe, Germany}
\date{\today}


\begin{abstract}
We review recent work on continuous quantum phase transitions in impurity
models, both with fermionic and bosonic baths -- these transitions are
interesting realizations of boundary critical phenomena at zero temperature.
The models with fermion bath are generalizations of the standard Kondo model,
with the common feature that Kondo screening of the localized spin
can be suppressed due to competing processes.
The models with boson bath are related to the spin-boson model of dissipative
two-level systems, where the interplay between
tunneling and friction results in multiple phases.
The competition inherent to all models can generate unstable fixed points associated
with quantum phase transitions, where the impurity properties undergo qualitative
changes.
Interestingly, certain impurity transitions feature both lower-critical and upper-critical
``dimensions'' and allow for epsilon-type expansions.
We present results for a number of observables, obtained by both analytical and
numerical renormalization group techniques, and make connections to experiments.
\end{abstract}


\maketitle


\section{Introduction}

Quantum mechanical systems can undergo zero-temperature phase
transitions upon variation of a non-thermal control parameter \cite{book,rop},
where order is destroyed solely by quantum fluctuations.
Quantum phase transitions occur as a result of competing ground
state phases, and can be classified into first-order and continuous
transitions.
The transition point of a continuous quantum phase transition,
the so-called quantum-critical point, is typically characterized
by a critical continuum of excitations, and can lead to
unconventional behavior -- such as non-trivial power laws
or non-Fermi liquid physics - over a wide range of the phase
diagram.

An interesting class of quantum phase transitions are so-called
{\em boundary} transitions where only the degrees of freedom of a subsystem
become critical.
In this paper we consider impurity transitions --
the impurity can be understood as a zero-dimensional boundary --
where the {\em impurity} contribution to the free energy
becomes singular at the quantum critical point.
Historically, the first transitions considered were the ones
in the dissipative spin-boson \cite{leggett,weiss} and the
anisotropic Kondo models \cite{yuval,hewson}.
Impurity quantum phase transitions require the
thermodynamic limit in the bath system, but are completely
independent of possible {\em bulk} phase transitions in the
bath.

In this paper we review recent work on quantum impurity models
undergoing zero-temperature phase transitions,
associated with intermediate-coupling fixed points.
These transitions can occur both in Kondo-type models with a fermionic bath
and in quantum dissipative models with a bosonic bath.
We shall show that these two classes are not fundamentally different,
and the transitions -- albeit in different universality classes -- can be
analyzed using similar techniques.
We elaborate on the existence of both lower-critical and upper-critical
``dimensions'' and the possibility to access critical behavior via
epsilon-type expansions.

All models have the general form
\begin{equation}
{\cal H} = {\cal H}_{\rm b} + {\cal H}_{\rm imp} \,.
\label{hgen}
\end{equation}
${\cal H}_{\rm b}$ contains the bulk degrees of freedom,
which generically are interacting;
however, under certain circumstances the self-interaction is irrelevant and
can be discarded from the outset.
${\cal H}_{\rm imp}$ contains the impurity degrees of freedom, e.g., one or more quantum spins,
together with their coupling to the bath, which typically is local in space.
Non-trivial quantum critical behavior in a model of the form (\ref{hgen})
obtains only if the thermodynamic limit for the bulk system ${\cal H}_{\rm b}$
is taken {\em before} the $T\to 0$ limit.

The paper is organized as follows:
In Sec.~\ref{sec:syst} we outline possible types of transitions which have
distinct finite-temperature crossover behavior, namely first-order,
second-order, and infinite-order transitions.
Issues related to the bath ${\cal H}_{\rm b}$ are discussed in Sec.~\ref{sec:bath},
in particular we emphasize that in certain cases bosonic baths cannot be
treated as Gaussian.
Sec.~\ref{sec:obs} introduces important observables, together with
their scaling behavior and related critical exponents.
The remainder of the paper deals with concrete model systems showing impurity
transitions.
We start with the most prominent example of intermediate-coupling impurity physics,
found in the fermionic multi-channel Kondo problem (Sec.~\ref{sec:2ck}).
Distinct non-trivial behavior can arise from a power-law variation of the
bath density of states at low energies --
this is described for the pseudogap Kondo model (Sec.~\ref{sec:pgk}),
the sub-ohmic spin-boson model (Sec.~\ref{sec:spb}), and
the Bose Kondo model (Sec.~\ref{sec:bk}).
Impurities with both fermionic and bosonic baths are subject of Sec.~\ref{sec:bfk},
whereas Secs.~\ref{sec:2imp} and \ref{sec:orb} deal with systems of coupled and
multi-orbital impurities, respectively.
Finally, Sec.~\ref{sec:qucl} is devoted to the quantum--classical mapping
commonly employed to discuss quantum phase transitions. We highlight
that this mapping is inapplicable for many situations encountered in
the context of quantum impurity models.
A discussion of general aspects will close the paper.
Most results quoted in this review have been obtained by
large-$N$, perturbative renormalization group (RG) and
Wilson's numerical renormalization group \cite{nrg,kri80} (NRG) methods;
we shall also mention some results from Bethe ansatz and conformal field theory
techniques.
Applications from various fields, e.g., magnetic impurities in non-metallic systems,
strongly interacting quantum dots, and noise in mesoscopic systems, will be
mentioned throughout the text.


\begin{widetext}

\begin{figure}[!t]
\epsfxsize=5.5in
\centerline{\epsffile{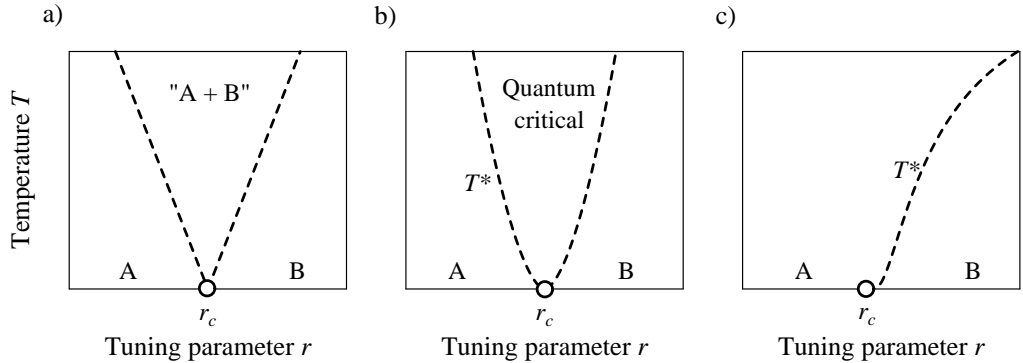}}
\caption{
Schematic finite-temperature phase diagrams for impurity quantum phase
transitions; A and B are the stable phases (e.g. the localized and delocalized
phases in the spin-boson model).
a) First-order transition, i.e., level crossing.
The finite-temperature properties are simply a thermodynamic mixture of A and B.
b) Second-order transition, with power laws.
The quantum-critical region, bounded by $T^\ast \propto |t|^{\nu}$
(where $t = (r-r_c)/r_c$ is the dimensionless measure of the distance
to the transition point),
is controlled by an unstable RG fixed point.
c) Infinite-order transition of Kosterlitz-Thouless type.
No unstable fixed point is present, and the leading thermodynamic behavior
displays only one crossover line. $T^\ast$ vanishes exponentially at
the transition point.
}
\label{fig:xover}
\end{figure}

\end{widetext}

\section{Transitions, critical dimensions, and finite-temperature crossovers}
\label{sec:syst}

Impurity quantum phase transitions can be continuous or
discontinuous (first-order); the class of continuous transitions
can be subdivided into transitions with power-law behavior
(typically second-order) and those with exponential behavior
(infinite-order transition, e.g., Kosterlitz-Thouless).
Quantum impurity models permit sharp boundary transitions only at $T=0$.
Depending on the type of transition, the finite-temperature crossovers
are different, see Fig.~\ref{fig:xover}.

A first-order transition is a simple level crossing in the system's
ground state: There exist two disconnected minima in the
energy landscape of the system, and the finite-$T$ properties above the
transition point are a simple thermodynamic mixture of the two phases.
A true quantum critical region is present for second-order transitions --
this will be the main focus of the present article.

Some of the examples presented below show second-order transitions
with exponents depending upon a continuous parameter, which specifies
the exponent of a low-energy power law in the bath density of states
which takes the role of a ``dimension'' --
this applies to the pseudogap Kondo, the spin-boson, and the
Bose-Fermi Kondo problems.
As a function of this ``dimension'', the line of second-order
transitions can terminate at a ``lower-critical dimension'',
with diverging correlation length exponent as this dimension is
approached.
Precisely at the lower-critical dimension the transition can turn
into one of Kosterlitz-Thouless type.
The pseudogap Kondo problem also features an ``upper-critical dimension''
where the interacting critical fixed point turns into a non-interacting
one. More precisely, above the upper-critical dimension the transition
becomes first-order with perturbative corrections, see Sec.~\ref{sec:pgk}


\section{Baths}
\label{sec:bath}

The baths appearing in quantum impurity models typically consist
of fermions or bosons; in the latter case, these can be either real
particles or quasiparticle collective excitations.
Of relevance for the impurity behavior is the local density of states
of the bath at the impurity location,
with an ultraviolet cutoff $\Lambda$ and a certain asymptotic
low-energy form. Most interesting here are gapless spectra with power-law
behavior at low energies, as will appear in many models below.

However, in a number of cases properties of the bath beyond its
local spectrum are important, and this must be discussed in the context
of the concrete physical model underlying the bath.
In particular, a free Bose field with a gapless spectrum is a delicate object
which can become unstable under infinitesimal perturbations (this should be
contrasted from a free Fermi field, which has a robust stability).
One instance of such an instability is the response of spin-1 bosons
to an applied magnetic field where the system is unstable to
arbitrarily large fluctuations.
This means that the self-interaction of the bath particles {\em cannot}
be neglected, and must be treated at an equal footing with coupling between
impurity and bath.
Technically, there is a non-trivial ``interference'' between the two couplings,
and the bath self-interaction significantly modifies the environment coupling to
the impurity.
It is thus not permissible to treat the environment as a Gaussian quantum noise,
and focus exclusively on the coupling to the impurity.
This applies e.g. to magnetic fluctuations at a bulk quantum critical point below
its upper-critical dimension,
as exemplified for the Bose Kondo model in Sec.~\ref{sec:bk}.

Conversely, in models with {\em non-interacting} bath particles
the environment is Gaussian, and progress can be made by integrating out the bath.
This results in a quantum model for the impurity only, with a (typically long-range)
self-interaction in imaginary time direction.
Such a model is effectively $(0+1)$-dimensional, and is believed to be
asymptotically equivalent to a one-dimensional classical (spin) model
by the virtue of the quantum--classical mapping.
A paradigmatic example is the $1/r^2$ Ising chain which is equivalent to both
the ohmic spin-boson model and the anisotropic Kondo model \cite{leggett}.
However, recent investigations of the sub-ohmic spin-boson model indicate
that the naive quantum--classical mapping can fail for certain types of long-range
interactions \cite{quclfail}, see Sec.~\ref{sec:qucl}.
For spin impurity models with higher symmetry [XY or SU(2)] the quantum--classical
mapping also fails; this can be traced back to the effect of the Berry phase
describing the spin dynamics, which has no classical analogue.


\section{Observables and scaling}
\label{sec:obs}

In this section we introduce a few observables which can be used to
characterize the phases and phase transitions discussed below.
The equations will be given for a situation where the impurity degree of
freedom is a single SU(2) spin $\bf \SSS$ of size $S$;
generalizations to other impurities are straightforward.
The models to be discussed display both ``screened'' and ``unscreened'' phases,
where a ``screened'' phase has a unique ground state with quenched impurity
degrees of freedom, whereas an ``unscreened'' phase shows a degenerate ground
state.

\subsection{Susceptibilities}

Canonical response functions can be derived using a local magnetic field
coupled to the impurity spin, by adding
\begin{eqnarray}
- H_{\text{imp},\alpha} {\SSS}_{\alpha}
\label{par2}
\end{eqnarray}
to the impurity Hamiltonian ${\cal H}_{\rm imp}$.
For a spinful bath, an external magnetic field can also be applied
to the degrees of freedom in ${\cal H}_{\rm b}$.
For conduction electrons $c_\sigma(x)$ the appropriate form is
\begin{eqnarray}
&&- H_{\text{u}\alpha}(x) (c^\dagger_\sigma \sigma_{\sigma\sigma'}^\alpha c_{\sigma'})(x)
\end{eqnarray}
The bulk field $H_{\text{u}}$ varies slowly as function of the space coordinate.

With these definitions, a spatially uniform field applied
to the whole system corresponds to $H_{\rm u} = H_{\rm imp} = H$.
Response functions can be defined from second derivatives of the thermodynamic
potential, $F = - T \ln Z$, in the standard way \cite{science,vbs}:
$\chi^{\alpha\alpha}_{\rm{u},\rm{u}}$ measures the bulk response to a field applied
to the bulk, $\chi^{\alpha\alpha}_{\rm{imp},\rm{imp}}$ is the impurity response to
a field applied to the impurity, and $\chi^{\alpha\alpha}_{\rm{u},\rm{imp}}$
is the cross-response of the bulk to an impurity field.
In the absence of global SU(2) symmetry, response functions for different
field directions have to be distinguished, and only some of them will be
singular at criticality.

The impurity contribution to the total susceptibility is defined as
\begin{eqnarray}
\chi_{\rm imp}(T)
= \chi_{\rm imp,imp} + 2 \chi_{\rm u,imp} + (\chi_{\rm u,u} - \chi_{\rm u,u}^{\rm bulk})
\,,
\end{eqnarray}
where $\chi_{\rm u,u}^{\rm bulk}$ is the susceptibility of the bulk system in
absence of the impurity.
For an unscreened impurity spin of size $S$ we expect
$\chi_{\rm imp}(T\to 0) = S(S+1)/(3T)$ in the low-temperature limit --
note that this unscreened moment will be spatially smeared out due to the
residual coupling to the bath.
A fully screened moment will be characterized by $T\chi_{\rm imp} = 0$.
In the presence of global SU(2) symmetry,
the susceptibility $\chi_{\rm imp}$ does {\em not} acquire an anomalous dimension
at criticality \cite{ss} (in contrast to $\chi_{\rm loc}$ below),
because it is a response function associated to the conserved quantity $S_{\rm tot}$.
Thus we expect a Curie law
\begin{equation}
\lim_{T\to 0} \chi_{\rm imp}(T) = \frac{{\cal C}_{\rm imp}}{T} \,,
\label{fract}
\end{equation}
where the prefactor ${\cal C}_{\rm imp}$ is in general a non-trivial universal constant
different from the free-impurity value $S(S+1)/3$.
Apparently, Eq.~(\ref{fract}) can be interpreted as the Curie response of a
fractional effective spin \cite{science} -- examples are e.g. found
in the pseudogap Kondo model (Sec.~\ref{sec:pgk}) and in the Bose Kondo model (Sec.~\ref{sec:bk})
below.

The local impurity susceptibility is given by
\begin{equation}
\chi_{\rm loc}(T) = \chi_{\rm imp,imp}  \,,
\label{chiloc}
\end{equation}
which is equivalent to the zero-frequency impurity
spin autocorrelation function.
In an unscreened phase we have $\chi_{\rm loc}\propto 1/T$ as $T\to 0$.
This Curie law defines a residual local moment $m_{\rm loc}$ at $T=0$,
which is the fraction of the total, freely fluctuating, moment of size $S$,
which remained localized at the impurity site:
\begin{equation}
\lim_{T \rightarrow 0} \chi_{\text{loc}}(T) = \frac{m_{\text{loc}}^2}{T} \,.
\label{mloc}
\end{equation}
A decoupled impurity has $m_{\text{loc}}^2={\cal C}_{\rm imp}=S(S+1)/3$,
but a finite coupling to the bath implies $m_{\text{loc}}^2 < {\cal C}_{\rm imp}$.
The quantity $m_{\rm loc}$ turns out to be a suitable order parameter \cite{insi}
for the phase transitions between an unscreened and a screened spin:
at a second-order transition it vanishes continuously as $t \to 0^-$.
Here, $t = (r-r_c)/r_c$ is the dimensionless measure of the distance
to criticality in terms of coupling constants,
with $t>0$ ($t<0$) placing the system into the (un)screened phase.
Thus, $T\chi_{\rm loc}$ is {\em not} pinned to the value of $S(S+1)/3$
for $t<0$ (in contrast to $T\chi_{\rm imp}$).

For models where the bulk does not carry spin, like the spin-boson model,
a {\em local} susceptibility can still be defined, and
$\chi^{zz}_{\rm loc}$ will be singular at criticality.

\subsection{Scaling ansatz and critical exponents}

A scaling ansatz for the impurity part of the free energy takes the form
\begin{equation}
F_{\rm imp} = T f(t T^{-1/v}, H_{\rm imp} T^{-b} )
\label{fscal}
\end{equation}
where $t$ measures the distance to criticality, and $H_{\rm imp}$
is the local field.
$\nu$ is the correlation length exponent which describes the vanishing of the energy scale
$T^\ast$, above which critical behavior is observed \cite{book}:
\begin{equation}
T^\ast \propto |t|^{\nu} .
\label{defnu}
\end{equation}
Note that there is no independent dynamical exponent $z$ for the
present $(0\!+\!1)$-dimensional models, formally $z=1$.
The ansatz (\ref{fscal}) assumes the fixed point to be interacting;
for a Gaussian fixed point the scaling function will also depend upon
dangerously irrelevant variables.

With the local magnetization
$M_{\rm loc} = \langle\SSS_z\rangle = -\partial F_{\rm imp}/\partial H_{\rm imp}$ --
note that $M_{\rm loc}(T\!\to\!0) = m_{\rm loc}$ (\ref{mloc}) --
and the corresponding susceptibility
$\chi_{\rm loc} = -\partial^2 F_{\rm imp}/(\partial H_{\rm imp})^2$
we can define critical exponents as usual:
\begin{eqnarray}
M_{\text{loc}}(t<0,T=0,H_{\rm imp}\to0)
&\propto& (-t)^{\beta}, \nonumber\\
\chi_{\text{loc}}(t>0,T=0) &\propto& t^{-\gamma},
\nonumber\\[-1.75ex]
\label{exponents} \\[-1.75ex]
M_{\text{loc}}(t=0,T=0) &\propto& | H_{\rm imp} |^{1/\delta}, \nonumber\\
\chi_{\text{loc}}(t=0,T) &\propto&
T^{-x}, \nonumber \\
\chi_{\text{loc}}''(t=0,T=0,\omega) &\propto&
|\omega|^{-y} {\rm sgn}(\omega). \nonumber
\end{eqnarray}
The last equation describes the dynamical scaling of the local susceptibility.

In the absence of a dangerously irrelevant variable, there are only two independent
exponents. The scaling form (\ref{fscal}) allows to derive hyperscaling relations:
\begin{eqnarray}
\beta = \gamma \frac{1-x}{2x},~~
2\beta + \gamma = \nu, ~~
\delta = \frac{1+x}{1-x} \,.
\end{eqnarray}
Furthermore, hyperscaling also implies $x=y$.
This is equivalent to so-called $\omega/T$ scaling in the dynamical
behavior -- for instance, the local dynamic susceptibility will obey
the full scaling form \cite{book}
\begin{equation}
\chi''_{\rm loc}(\omega,T)
= \frac {{\cal B}_1} {\omega^{1-\eta_{\chi}}} \, \Phi_1 \!\left(\frac{\omega}{T}, \frac{T^{1/\nu}}{t}\right)
\label{scalchi1}
\end{equation}
which describes critical local-moment fluctuations,
and the local static susceptibility follows
\begin{equation}
\chi_{\rm loc}(T)
= \frac {{\cal B}_2} {T^{1-\eta_{\chi}}} \, \Phi_2 \!\left(\frac{T^{1/\nu}}{t}\right) \,.
\label{scalchi2}
\end{equation}
which reduces to the form quoted in (\ref{exponents}) at criticality, $t=0$.
Here, $\eta_{\chi} = 1-x$ is a universal anomalous exponent,
and $\Phi_{1,2}$ are universal crossover functions (for the specific critical
fixed point), whereas
${\cal B}_{1,2}$ are non-universal prefactors.

\subsection{Impurity entropy}

In general, zero-temperature impurity critical points
can show a non-trivial residual entropy [contrary
to bulk quantum critical points where the entropy usually vanishes
with a power law, $S(T) \propto T^y$].
The impurity entropy is evaluated as entropy of the system with
impurity minus entropy of the bath alone.
The stable phases usually have impurity entropies of the form
$S_{\rm imp}(T\to 0) = \ln g$ where is the integer ground state degeneracy,
e.g., $g=1$ for a Kondo-screened impurity and $g=2S+1$ for an unscreened
spin of size $S$.
At a second-order transition $g$ can take fractional values;
here it is again important that the thermodynamic limit in the bath
is taken {\em before} the $T\to 0$ limit.

Thermodynamic stability requires that the total entropy of a system decreases
upon decreasing temperature, $\partial_T S(T) > 0$.
This raises the question of whether the impurity part of the entropy, $S_{\rm imp}$,
has to decrease under RG flow (which is equivalent to decreasing $T$).
The so-called $g$-theorem \cite{gtheorem} provides a proof of this conjecture
for systems with short-ranged interactions;
for most quantum impurity problems this appears to apply.
Interestingly, both the pseudogap Kondo model and the spin-boson model provide
explicit counter-examples, see Secs.~\ref{sec:pgk} and \ref{sec:spb}, with
details in Refs.~\onlinecite{MVLF} and \onlinecite{bosnrg}.
(For another counter-example see Ref.~\onlinecite{uphill}.)

On the technical side, we
note that the impurity part of the thermodynamic potential, $F_{\rm imp}$, will
usually diverge with the ultraviolet cutoff $\Lambda$ of the bath.
Thus we have $F_{\rm imp} = E_{\rm imp} - T S_{\rm imp}$,
where $E_{\rm imp}$ is the non-universal (cutoff-dependent) impurity contribution
to the ground-state energy.
However, the impurity entropy $S_{\rm imp}$ is fully universal in the
limit $T/\Lambda\to 0$.

\subsection{$T$ matrix}

Depending on the type of the impurity, further observables can be used
to characterize the behavior near criticality.
In an Anderson or Kondo model the conduction electron $T$ matrix describes
the scattering of the conduction electrons off the impurity.
For an Anderson model, the $T$ matrix is (up to a prefactor) equivalent to
the full impurity electron Green's function.
As with the local susceptibility, we expect a scaling form of the
$T$ matrix spectral density near the intermediate-coupling fixed points
similar to Eq.~(\ref{scalchi1}).
In particular, at criticality a power law occurs:
\begin{equation}
T(\omega) \propto \frac{1}{\omega^{1-\eta_T}} \,.
\label{tmatpower}
\end{equation}
We will evaluate the anomalous exponent $\eta_T$ for certain models below.

Notably, the $T$ matrix can be directly observed in experiments on Kondo impurities,
due to recent advances in low-temperature scanning tunneling microscopy.
This has been demonstrated for magnetic ions on metal surfaces \cite{cocu,cocumulti},
and also with impurity moments in high-temperature superconductors \cite{seamus}.
Further, both metallic and semiconductor quantum dots in the Coulomb blockade regime
can show Kondo physics \cite{kondodot1,kondodot2},
and transport experiments through the dot then probe the local
spectral function via the differential conductance.

\subsection{Local non-Fermi liquid behavior}

Magnetic impurities in metals which show the standard Kondo effect
can be described in terms of local Fermi liquid physics, i.e.,
the imaginary part of the conduction electron self-energy
follows $\omega^2$ at low energies, the impurity contribution to
the entropy is $\gamma T$ etc.
In general, the impurity influence
on the bulk system can be captured by a potential scatterer with
energy-dependent phase shift.

In contrast, the intermediate-coupling fixed points to be described below
are associated with local non-Fermi liquid behavior, and the above
concepts do not apply.
This is immediately clear e.g. for the impurity entropy which can be
finite as $T\to 0$.
A paradigmatic and well-studied example for local non-Fermi liquid behavior is
found in the two-channel Kondo effect (Sec.~\ref{sec:2ck}).


\section{Multi-channel Kondo model}
\label{sec:2ck}

Many of the models to be discussed in this paper are built from magnetic
moments which can show the Kondo effect \cite{hewson}.
Originally, this effect describes the behavior of localized
magnetic impurities in metals. The relevant microscopic models
are the Kondo model and the single-impurity Anderson
model.
In the standard case (i.e. a single magnetic
impurity with spin 1/2 coupled to a single conduction
band with a finite density of states (DOS) near the
Fermi level) screening of the magnetic
moment occurs below a temperature scale $\TK$.
The screening is associated with the flow to strong coupling
of the effective interaction between impurity and host fermions,
i.e., perturbation theory diverges on the scale $\TK$.
The Kondo temperature $\TK$
depends exponentially on the model parameters.
For $T \ll \TK$ the system behaves as a local Fermi liquid.

Kondo screening is strongly modified if
two or more fermionic screening channels compete and the moment
is then overcompensated.
Nozi\`eres and Blandin \cite{Noz80} proposed a
two-channel generalization of the Kondo model, which shows
overscreening associated with an intermediate-coupling fixed point
and non-Fermi liquid behavior
in various thermodynamic and transport properties.

The Hamiltonian for a multi-channel Kondo model can be written as
${\cal H} = {\cal H}_{\rm MK} + \sum_i {\cal H}_{{\rm b},i}^{(\rm f)}$, with
\begin{eqnarray}
\label{mkm}
{\cal H}_{\rm MK} &=& \sum_{i=1}^K J_i \, {\bf \SSS} \cdot {\bf s}_i(0) \\
{\cal H}_{{\rm b},i}^{(\rm f)} &=& \int_{-\Lambda}^{\Lambda} dk \,
  k\, c_{k \sigma i}^\dagger c_{k\sigma i}
\nonumber
\end{eqnarray}
where $i=1, \ldots, K$ is the channel index, and
summation over repeated spin indices $\sigma$ is implied.
The fermionic baths ${\cal H}_{\rm b}^{(\rm f)}$ are represented by one-dimensional
chiral fermions $c_{k}$, with a density of states $\rho_0$ at the Fermi level
and an ultraviolet (UV) cutoff $\Lambda$.
The impurity spin $\bf \SSS$ is coupled to the conduction electron
spins at site $0$, $s_{\alpha i}(0) = c_{\sigma i}^\dagger(0) \sigma_{\sigma\sigma'}^\alpha c_{\sigma' i}(0) / 2$
with $c_{\sigma i}(0) = \int d k k \, c_{k \sigma i}$,
and $\sigma^\alpha$ is the vector of Pauli matrices.

\begin{figure}[!t]
\epsfxsize=2.3in
\centerline{\epsffile{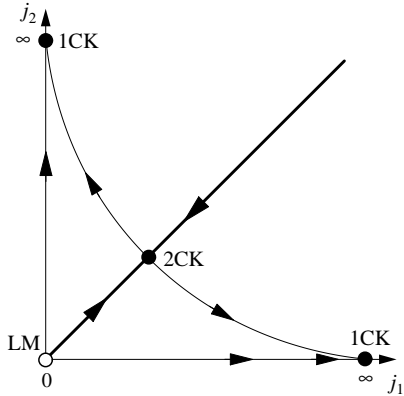}}
\caption{
RG flow diagram for the two-channel Kondo model
where the two axes denote the renormalized couplings $j_1,j_2$
in the two channels, with $j_i = \rho_0 J_i$ initially.
LM is the local-moment fixed point of a decoupled impurity,
1CK is the strong-coupling fixed point of the single-channel Kondo effect,
and 2CK is the intermediate-coupling two-channel fixed point.
}
\label{fig:2ckflow}
\end{figure}

In general, overscreening occurs for any number of channels $K>1$
coupled to a spin 1/2.
It does not require fine-tuning of the Kondo coupling, provided that the
couplings $J_i$ in different channels are equal, i.e., if the
system obeys SU(2)$_{\rm spin} \times$ SU($K$)$_{\rm channel}$ symmmetry \cite{CZ}.
A quantum phase transition can be driven by a channel asymmetry,
with the RG flow diagram for the two-channel case shown in Fig.~\ref{fig:2ckflow}.
One of the two equivalent strong-coupling Fermi liquid fixed points is reached
for unequal couplings $J_1 \neq J_2$, whereas for $J_1 = J_2$ the system
flows to the two-channel intermediate-coupling fixed point.
Thus, the two-channel fixed point acts as critical fixed point separating the
two Fermi liquid phases.

As an aside we note that the two-channel Anderson model has recently
been shown to have even richer behavior \cite{2caim}:
it displays a line of non-Fermi liquid fixed points which continuously connects
the two-channel spin and two-channel charge Kondo effects.

\subsection{RG}

The existence of an intermediate-coupling fixed point
for a large number of screening channels $K$
can be seen from the RG flow equation for the
dimensionless Kondo coupling, $j = j_1 = j_2$ ($j_i = \rho_0 J_i$ initially),
in the channel-symmetric case
\begin{equation}
\beta(j) = -j^2 + \frac{K}{2} j^3
\end{equation}
to two-loop order.
It predicts an unstable fixed point at $j^\ast = 2/K$,
which indeed exists for all $K\geq 2$ and is perturbatively
accessible in the limit of large channel number,
$K\gg 1$ \cite{Noz80,olivier}.

\subsection{Exact solution}

The multi-channel Kondo model can be exactly
solved by means of the Bethe ansatz \cite{bethe2ck}.
In addition, many of the low-energy properties of the two-channel Kondo and
related models have been successfully studied using conformal field
theory techniques \cite{AL}.
This allows to calculate the leading temperature dependencies of
physical observables, like entropy, suceptibility, and
also dynamic multipoint correlation functions which are not directly
accessible to Bethe ansatz methods.
For a more detailed description we refer the reader to the
literature \cite{AL,affleck2ck,fuji2ck}.

\subsection{Observables}

The anomalous low-temperature properties at the multi-channel
non-Fermi liquid fixed point are controlled by the leading irrelevant
coupling (see below), which has scaling dimension $y = 2/(2+K)$.
One finds for the impurity susceptibility and the specific heat ratio
\begin{eqnarray}
\chi_{\rm imp} &\propto& (T/\TK)^{2y-1} \,, \nonumber\\
\gamma = C_{\rm imp}/T &\propto& (T/\TK)^{2y-1} \,.
\end{eqnarray}
In the two-channel case, $K=2$, logarithmic corrections occur, leading to
\begin{eqnarray}
\chi_{\rm imp} &\propto& \ln(\TK/T) \,, \nonumber\\
\gamma = C_{\rm imp}/T &\propto& \ln(\TK/T) \,.
\end{eqnarray}
For $K=2$ the residual entropy is $S_{\rm imp} = \frac{1}{2} \ln 2$,
and an anomalous Wilson ratio $R=\chi/\gamma=8/3$ appears, in contrast
to the result for the standard Kondo model $R=8/4=2$.

We caution the reader that the quoted behavior of $\chi_{\rm imp}$
violates the naive scaling expectation $\chi_{\rm imp} \propto 1/T$ (\ref{fract}).
Even though the low-energy physics is controlled by an intermediate-coupling fixed point,
a ``compensation'' effect \cite{barzykin} causes most thermodynamic response functions
to vanish in the naive scaling limit, and the leading low- temperature behavior
is exposed only upon considering corrections to scaling \cite{AL,olivier}.
For this reason $\chi_{\rm imp}$ and $\chi_{\rm loc}$ follow the same anomalous
behavior, in contrast to the pseudogap Kondo and Bose Kondo models, described
in Secs.~\ref{sec:pgk} and \ref{sec:bk}.

\subsection{Applications}

Experimental realizations of two-channel Kondo physics have been discussed early on
in the context of certain rare-earth compounds. The idea is that in
materials like UBe$_{13}$ orbital (i.e. quadrupolar) degrees of freedom can be
screened by the interaction with conduction electrons, and the electron
spin $\sigma=\uparrow,\downarrow$ provides the two channels required
for overscreening. This proposal has been controversial, and we refer the
reader to a recent review \cite{CZ} for details.

As the Kondo effect of a single spin can be conveniently investigated in
transport experiments through quantum dots in the Coulomb blockade regime \cite{kondodot1,kondodot2},
various proposals for the realization of two-channel Kondo physics in
quantum-dot devices have been put forward.
The difficulties for the spin two-channel Kondo effect are that the
screening channels need to be independent
(i.e. no mixing due to cotunneling between the bands) and the couplings to the
channels need to be equal.
Provided that cotunneling is suppressed, the channel asymmetry may
be tuned by suitable gate voltages, leading to a quantum phase transition
as illustrated in Fig.~\ref{fig:2ckflow}; the transport signatures of
this phase transition have been detailed in Ref.~\onlinecite{glazman}.
Versions of the two-channel Kondo effect utilizing {\em charge} degrees of freedom
in quantum dots \cite{matveev} naturally satisfy the above requirements,
as the two independent channels are provided by the electron spin.
Here, the problem is to reach temperatures sufficiently below $\TK$.

A recent proposal by Oreg and Goldhaber-Gordon \cite{oreg} utilizes Coulomb blockade
physics in the leads, formed by metallic islands, to suppress cotunneling between
different leads. Subsequently it has been shown \cite{uphill} that a tiny charging energy
in those leads is sufficient to stabilize multi-channel physics as $T\to 0$.
In such a situation, a single-channel Kondo resonance develops at intermediate temperature,
which is destroyed at lowest $T$, i.e.,
the system crosses over from single- to multi-channel physics (!) upon lowering temperature.
This ``reversed'' renormalization group flow is opposite to the standard behavior
in a multi-channel model with small channel anisotropy, Fig.~\ref{fig:2ckflow}.
It contradicts the conventional wisdom that single-channel Kondo physics
cannot be destroyed by small perturbations -- the physical reason is the
long-range nature of the Coulomb charging energy \cite{uphill}.

We note that to date experimental efforts in realizing the two-channel Kondo effect
in quantum dots have remained unsuccessful.


\section{Pseudogap Kondo and Anderson models}
\label{sec:pgk}

A straightforward possibility to suppress Kondo screening in a single-channel
situation is to reduce the electron bath density of states at the Fermi level
to zero.
The absence of low-energy states prevents screening for small
Kondo couplings.
If the bath gap arises from host superconductivity this can
be interpreted as competition between Kondo singlet formation and
Cooper pairing.

Two different types of gaps have to be distinguished:
so-called hard-gap and pseudogap (soft-gap) systems.
In the hard-gap case, the DOS $\rho(\varepsilon)$ is zero in a
finite energy interval around the Fermi level -- this is e.g. realized in
a conventional superconductor.
The resulting impurity transition between a local-moment (LM) phase without Kondo
screening, realized at small Kondo coupling $J$, and a screened
strong-coupling (SC) phase, reached for large $J$, is of first order,
and it occurs only in the presence of particle-hole (p-h) asymmetry~\cite{hardgap}.
In the p-h symmetric case, the local-moment state
persists for arbitrary values of the coupling.

The pseudogap case, first considered by Withoff and Fradkin \cite{withoff},
corresponds to a bath with
$\rho(\varepsilon) \propto |\varepsilon|^r$ ($r>0$), i.e.,
the DOS is zero only {\em at} the Fermi level.
The corresponding Kondo and single-impurity Anderson models interpolate between
the metallic case ($r=0$) and the hard-gap case ($r\to\infty$).
The pseudogap case $0<r<\infty$ leads to a very rich behaviour,
in particular to a continuous transition
between a local-moment and a strong-coupling phase \cite{withoff,cassa,tolya2,GBI}
Central to our discussion will be the single-impurity
Anderson model with a pseudogap host density of states,
${\cal H} = {\cal H}_{\rm A} + {\cal H}_{\rm b}^{(\rm f)}$:
\begin{eqnarray}
\label{aim}
{\cal H}_{\rm A} &=& \epsfb f_\sigma^\dagger f_\sigma + \Ub n_{f\uparrow} n_{f\downarrow}
  + \hybb \left(f_\sigma^\dagger c_\sigma(0) + {\rm h.c.}\right),
  \\
{\cal H}_{\rm b}^{(\rm f)} \!\! &=& \int_{-\Lambda}^{\Lambda} dk\,|k|^r \,
  k c_{k\sigma}^\dagger c_{k\sigma}
\nonumber
\end{eqnarray}
with notations as in Sec.~\ref{sec:2ck}, and the bath is now represented
by linearly dispersing fermions in $(1\!+\!r)$ dimensions.
In the Kondo limit of the Anderson model charge fluctuations are frozen out,
and the impurity site is mainly singly occupied.
Via Schrieffer-Wolff transformation one obtains the standard Kondo model,
${\cal H} = {\cal H}_{\rm K} + {\cal H}_{\rm b}^{(\rm f)}$, with
\begin{equation}
\label{km}
{\cal H}_{\rm K} = \Jb {\bf S} \cdot {\bf s}(0) \,.
\label{pgk}
\end{equation}
The Kondo coupling is related to the parameters of the Anderson model (\ref{aim})
through:
\begin{equation}
\Jb = 2 \hybb^2 \bigg(\frac{1}{|\epsfb|} + \frac{1}{|\Ub+\epsfb|}\bigg) \,.
\label{j-swo}
\end{equation}
The Kondo limit is reached by taking $\Ub\to\infty$, $\epsfb\to -\infty$,
$\hybb\to\infty$, keeping $\Jb$ fixed.
In the absence of p-h symmetry the Schrieffer-Wolff transformation
also generates a potential scattering term in the effective Kondo
model \cite{hewson}.

\begin{figure}[!t]
\epsfxsize=2.9in
\centerline{\epsffile{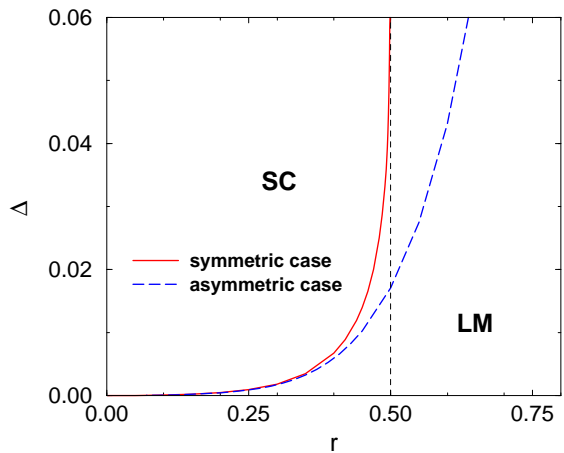}}
\caption{
$T=0$ phase diagram for the pseudogap Anderson model
in the p-h symmetric case (solid line, $U_0=10^{-3}$,
            $\varepsilon_0 = -0.5 \cdot 10^{-3}$, conduction band
            cutoff $\Lambda = 1$) and the p-h asymmetric case
            (dashed line,  $\varepsilon_0 = -0.4 \cdot 10^{-3}$);
            $\Delta$ measures the hybridization strength,
            $\pi V_0^2 \rho(\omega) = \Delta |\omega|^r$.
The lines of critical points separate
the strong coupling phase SC ($\Delta>\Delta_{\rm c}$)
from the local-moment phase LM ($\Delta<\Delta_{\rm c}$).
After Ref.~\onlinecite{bullapg}.
}
\label{fig:nrgpd}
\end{figure}

\subsection{Pseudogap Anderson model: Phase diagram and NRG}

Fig.~\ref{fig:nrgpd} shows a typical phase diagram for the pseudogap
Anderson model \cite{GBI,bullapg}.
In the p-h symmetric case (solid) the critical coupling $\Delta$,
measuring the hybridization between band electrons and local moment,
diverges at $r=1/2$, and no screening occurs for $r>1/2$.
No divergence occurs for p-h asymmetry (dashed).
Importantly, the strong-coupling phases in the p-h symmetric and
asymmetric situations belong to different low-energy fixed points,
in the following denoted by SSC and ASC, respectively.

We now briefly describe the stable phases;
NRG calculations have established that their low-energy properties are
identical for the pseudogap Anderson and Kondo models \cite{GBI}.
Due to the power-law conduction band DOS, already the stable LM
and SSC/ASC fixed points show unconventional behavior \cite{GBI,bullapg}.
The LM phase has the properties of a free spin 1/2
with residual entropy $S_{\rm imp}=\ln 2$ and
low-temperature impurity susceptibility $\chi_{\rm imp}=1/(4T)$,
but the leading corrections show $r$-dependent power laws.
The p-h symmetric SSC fixed point has very unusual properties,
namely $S_{\rm imp}=2 r \ln 2$, $\chi_{\rm imp}=r/(8 T)$
for $0<r<1/2$.
In contrast, the p-h asymmetric ASC fixed point simply displays
a completely screened moment, $S_{\rm imp}= T\chi_{\rm imp}=0$.
The impurity spectral function follows a $\omega^r$ power law
at both the LM and the ASC fixed point, whereas it
diverges as $\omega^{-r}$ at the SSC fixed point --
this ``peak'' can be viewed as a generalization of the Kondo resonance in
the standard case ($r=0$), and scaling of this peak is observed upon
approaching the SSC-LM phase boundary \cite{bullapg,David}.

Let us turn to the critical fixed points.
Early work employed a weak-coupling RG based on an expansion in the
Kondo coupling, see next subsection. It turns out that the applicability of this
approach is restricted to small $r$.
Interestingly, the NRG studies \cite{GBI} showed that the
fixed-point structure changes at $r=r^\ast\approx 0.375$
and also at $r\!=\!\frac{1}{2}$, rendering the interesting case of $r\!=\!1$
inaccessible from weak coupling.
For $r<r^\ast$ the p-h symmetric and asymmetric models reach the same critical
fixed point (SCR) which is p-h symmetric.
In contrast, for $r>r^\ast$ a p-h asymmetric critical fixed point (ACR) exists,
which then controls the phase transition in the p-h asymmetric model.
The SCR fixed point ceases to exist for $r\geq 1/2$.
In addition, the critical fluctuations in the p-h asymmetric case change their character
at $r\!=\!1$: whereas for $r\!<\!1$ the exponents take non-trivial $r$-dependent values
and obey hyperscaling, exponents are trivial for $r\!>\!1$ and hyperscaling is
violated \cite{GBI,insi}.

\begin{figure}
\epsfxsize=3.1in
\centerline{\epsffile{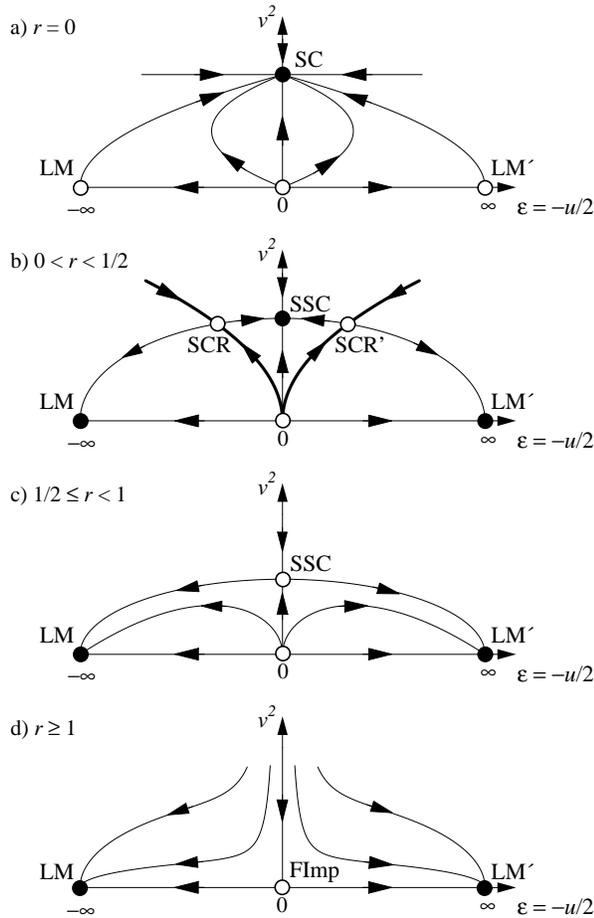}}
\caption{
Schematic RG flow diagrams for the particle-hole symmetric single-impurity
Anderson model with a pseudogap DOS, $\rho(\omega)\propto|\omega|^r$.
The horizontal axis denotes the renormalized on-site level energy $\epsf$
(related to the on-site repulsion $\U$ by $\U = -2 \epsf$),
the vertical axis is the renormalized hybridization $\hyb$.
The thick lines correspond to continuous boundary phase transitions;
the full (open) circles are stable (unstable) fixed points,
for details see text.
a) $r\!=\!0$, i.e., the familiar metallic case. For any finite $\hyb$
the flow is towards the strong-coupling fixed point (SC), describing Kondo screening.
b) $0\!<\!r\!<\!1/2$: The local-moment fixed point (LM) is stable,
and the transition to symmetric strong coupling (SSC) is controlled by the
SCR fixed point.
For $r\to 0$, SCR approaches LM, and the critical behavior at SCR is
accessible via an expansion in the Kondo coupling $j$.
In contrast, for $r\to 1/2$, SCR approaches SSC, and the critical behavior can be
accessed by expanding in the deviation from SCR, i.e., in $\epsf=-\U/2$.
c) $1/2\!\leq\!r\!<\!1$: $\hyb$ is still relevant at $\U=0$. However, SSC is now
unstable w.r.t. finite $\U$.
At finite $\hyb$, the transition between the two stable fixed points
LM and LM' is controlled by SSC (which is now a critical fixed point!).
d) $r\!\geq\!1$: $\hyb$ is irrelevant, and the only transition is a level crossing
(with perturbative corrections) occurring at $\hyb=\U=0$, i.e., at the free-impurity
fixed point (FImp).
After Ref.~\protect\onlinecite{MVLF}.
}
\label{fig:flowsym}
\end{figure}

Analytical understanding came with RG expansions applied to
the pseudogap Anderson model \cite{MVLF} which allow to formulate
the universal critical theories for $r$ near 1/2 and near 1.
In the p-h symmetric case the line of non-trivial phase transitions terminates
at {\em two lower-critical dimensions} (!), $r=0$ and $r=1/2$.
The following two subsections describe two RG expansions, with
small parameters $r$ and $(1/2-r)$ respectively,
which access the same critical fixed point (SCR) and are expected to match.
In the p-h asymmetric case an expansion can be
done in the hybridization around the valence-fluctuation point of the
Anderson model.
Bare perturbation theory is sufficient for all $r>1$;
for $r<1$ a perturbative RG procedure is required to calculate critical properties,
with the expansion being controlled in the small parameter $(1-r)$.
In particular, this identifies $r=1$ as the upper-critical dimension of
the (asymmetric) pseudogap Kondo problem, and consequently observables acquire
logarithmic corrections for $r=1$.
The full RG flow diagrams in the variables of the Anderson model (\ref{aim})
are displayed in Figs.~\ref{fig:flowsym} and \ref{fig:flowinfu} for
the p-h symmetric and asymmetric cases, respectively.
The properties near criticality will be discussed in detail in
Sec.~\ref{sec:pgobs} below.

\subsection{Pseudogap Kondo model: RG near $r=0$}
\label{sec:weak}

For small values of the DOS exponent $r$, the phase transition in the
pseudogap Kondo model can be accessed from the weak-coupling limit,
using a generalization of Anderson's poor man's scaling.
Power counting about the local moment fixed point (LM) shows that
the Kondo coupling $J$ is marginal at the lower-critical dimension $r=0$.
The flow of the renormalized Kondo coupling $j$ is given by the beta function
\begin{equation}
\label{betaj}
\beta(j) = r j - j^2 \,.
\end{equation}
For $r>0$ there is a stable fixed point at $j^\ast = 0$
corresponding to the local-moment phase (LM).
An unstable fixed point, controlling the transition
to the strong-coupling phase, exists at $j^\ast = r$,
and the critical properties can be determined in a double expansion in
$r$ and $j$.
P-h asymmetry is irrelevant, i.e., a potential scattering
terms scales to zero according to $\beta(v) = r v$,
thus the above expansion captures the p-h symmetric
critical fixed point (SCR).

\subsection{Pseudogap Anderson model: RG near $r=1/2$}
\label{sec:sym}

For $r$ near $1/2$ the p-h symmetric critical fixed point moves
to strong Kondo coupling,
and the language of the p-h symmetric Anderson model \cite{MVLF} becomes
more appropriate.
Expanding in the renormalized hybridization $v$ and on-site interaction $u$,
the flow equations
\begin{eqnarray}
\label{betau}
&&\beta(v) = - \frac{1-r}{2} \, v +  v^3 \,, \nonumber\\
&&\beta(u) = (1-2r)\, u -  \frac{3(\pi-2 \ln 4)}{\pi^2}\, u^3
\end{eqnarray}
capture various fixed points;
notably the RG equation for $v$ is {\em exact} to all orders \cite{MVLF}.
The RG flow is depicted in Fig.~\ref{fig:flowsym}.

For $r<1/2$ a stable fixed point is at
${v^\ast}^2 = (1-r) / 2$, $u^\ast=0$, which represents nothing but the physics of the
non-interacting resonant level model with a power-law density of states --
interestingly, the impurity spin is not fully screened for $r>0$, and
the residual entropy is $2r\ln 2$.
This fixed point is identical with the symmetric strong-coupling fixed point (SSC)
of Gonzalez-Buxton and Ingersent\cite{GBI}; it becomes unstable for
$r>1/2$.
Additionally, for $r<1/2$ there is a pair of critical fixed points (SCR,SCR') located at
${v^\ast}^2 = (1-r) / 2$,
${u^\ast}^2 = \pi^2 (1-2r) / [3(\pi-2\ln 4)]$.
These describe the transition between an unscreened (spin or charge) moment phase and
the symmetric strong-coupling (i.e. screened) phase.
Importantly, both (\ref{betaj}) and (\ref{betau}) capture the same critical
fixed point.
Using (\ref{betau}), the expansion is performed around the strong-coupling fixed point (SSC),
and is valid for $u^\ast\ll 1$, i.e., for $1/2-r \ll 1$.

\subsection{Pseudogap Anderson model: RG near $r=1$}
\label{sec:infu}

\begin{figure}
\epsfxsize=3in
\centerline{\epsffile{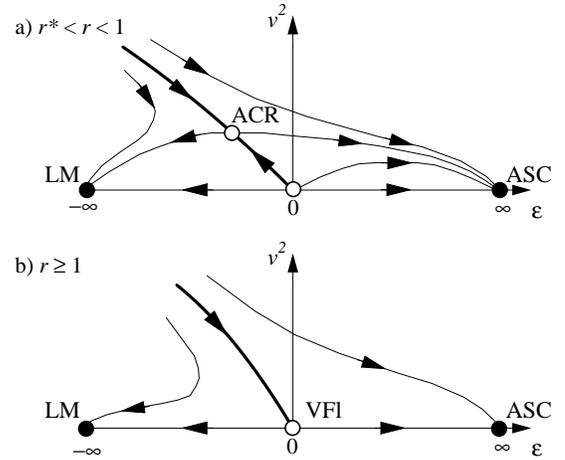}}
\caption{
Schematic RG flow diagrams for the maximally particle-hole asymmetric
pseudogap Anderson impurity model.
The horizontal axis denotes the on-site impurity energy, $\epsf$,
the vertical axis is the fermionic coupling $\hyb$,
the bare on-site repulsion is fixed at $\U_0=\infty$.
The symbols are as in Fig.~\protect\ref{fig:flowsym}.
a) $r^\ast<r\!<\!1$: $\hyb$ is relevant, and the transition is controlled by an interacting
fixed point (ACR). 
As $r\to r^\ast \approx 0.375$,
p-h symmetry at the critical fixed point is dynamically restored,
and ACR merges into the SCR fixed point of Fig.~\protect\ref{fig:flowsym} -- this
cannot be described using the RG of Sec.~\protect\ref{sec:infu}.
In the metallic $r=0$ situation,
the flow from any point with $\hyb\neq 0$ is towards the screened singlet fixed
point with $\epsf=\infty$.
b) $r\!\geq\!1$: $\hyb$ is irrelevant, and the transition is a level crossing
with perturbative corrections, occuring at $\hyb=\epsf=0$, i.e., the
valence-fluctuation fixed point (VFl).
After Ref.~\protect\onlinecite{MVLF}.
}
\label{fig:flowinfu}
\end{figure}

In the p-h asymmetric case a phase transition occurs for all $r>0$;
here the strong-coupling phase is maximally p-h asymmetric and corresponds to a fully
screened spin for all $r$.
Remarkably, for $0<r< r^{\star}\approx 0.375$ p-h symmetry is dynamically restored
at the phase transition, and the critical properties are described by the expansions discussed
above.
For $r>r^{\star}$ there exists an additional critical fixed point (ACR) with finite p-h asymmetry,
and its properties are best discussed using an infinite-$U_0$ Anderson model \cite{MVLF}.
Expanding around the $\epsilon=0$, $v=0$ limit, one finds the
flow equations
\begin{eqnarray}
&&\beta(v) = - \frac{1-r}{2} v + \frac{3}{2} v^3 \,, \nonumber \\
&&\beta(\epsilon) = - \epsilon + 3 v^2 \epsilon - v^2 \,,
\label{beta}
\end{eqnarray}
with a critical fixed point
${v^\ast}^2 = (1-r)/3$, $\epsilon^\ast = - (1-r)/3$ for $r<1$, and
$v^\ast=\epsilon^\ast=0$ for $r>1$.
The complete RG flow diagrams are in Fig.~\ref{fig:flowinfu}.

Note that the flow is very similar to the one of a $\phi^4$ model,
with $(1-r)$ playing the role of $(4-d)$, as
the Anderson model hybridization is marginal at $r=1$.
Clearly, the critical fixed point is interacting for $r<1$ (the analogue of the
Wilson-Fisher fixed point), whereas for $r>1$ we have a level crossing with
perturbative corrections (VFl, the analogue of the Gaussian fixed point).
This justifies to identify $r=1$ with the upper-critical dimension of the
problem.
RG analysis has shown that all fixed points presented above are stable
with respect to spin anisotropies.

\subsection{Pseudogap Kondo model: large-$N$}

A large-$N$ theory can be constructed by generalizing the spin symmetry of
both quantum impurity and conduction electrons from SU(2) to SU($N$).
An antisymmetric representation of the impurity spin uses
auxiliary fermions $f_\alpha$ ($\alpha=1, \ldots, N$):
\begin{equation}
S_{\alpha\beta} = f_\alpha^\dagger f_\beta - q \delta_{\alpha\beta}
\label{imprep}
\end{equation}
with a constraint
$\sum_\alpha f_\alpha^\dagger f_\alpha = qN$ enforced by a chemical potential $\lam$;
the value $q=1/2$ corresponds to p-h symmetry and will be used
in the following.

The single-channel Kondo model takes the form
\begin{equation}
{\cal H}_{\rm K} = -\frac{\Jb}{N} f^{\dagger}_{\alpha}
c_{\alpha} (0) c_{\beta}^{\dagger} (0) f_{\beta}
+ \lam f^{\dagger}_{\alpha} f_\alpha \,.
\end{equation}
In the limit of $N\!\to\!\infty$ the action is dominated by a static
saddle point where the field $b$ conjugate to
$f^{\dagger}_{\alpha}c_{\alpha} (0)$ acquires an expectation value.
We arrive at the non-interacting model
\begin{equation}
{\cal H}_{\rm K} =
- \left(b \, f^{\dagger}_{\alpha} c_{\alpha}(0) + {\rm h.c.}\right)
+ \lam f^{\dagger}_{\alpha} f_\alpha
+ \frac{N b^2}{\Jb}
\label{hbil}
\end{equation}
and the self-consistency equation
\begin{equation}
b = \frac{\Jb}{N} \sum_\beta \langle c_{\beta}^{\dagger} (0) f_{\beta} \rangle \,.
\label{beq}
\end{equation}
Eqs.~(\ref{hbil},\ref{beq}) constitute the well-known slave-boson
mean-field approximation. Non-zero $b$ signals Kondo screening.
If $b\neq 0$ at $T=0$ then $b$ will vanish continuously as a
temperature $\TK$ is approached from below -- this defines the Kondo
temperature. Note that the sharp finite-temperature transition is
an artifact of the large-$N$ limit.

For a metallic density of states, $r=0$, the slave-boson method yields
the correct exponential dependence of $\TK$ on $\Jb$.
For $r>0$ a $T=0$ transition at a finite $\Jb$ is predicted.
However, the critical properties of this transition are not
reproduced correctly; in particular at $r=1$ the slave-boson
method yields an essential singularity of $\TK$ near the transition
point\cite{cassa,tolya2} (instead of the linear behavior, see above).

An interesting alternative to the single-channel large-$N$ limit
is a multi-channel version, where the impurity is coupled to $K$
screening channels \cite{olivier}.
The Hamiltonian then reads
\begin{equation}
{\cal H}_{\rm MK} = -\frac{\Jb}{N} f^{\dagger}_{\alpha}
c_{\alpha i} (0) c_{\beta i}^{\dagger} (0) f_{\beta}
+ \lam f^{\dagger}_{\alpha} f_\alpha
\end{equation}
where $i=1, \ldots, K$ is the channel index.
Taking the $N\!\to\!\infty$ limit with $K=\gamma N$ yields a {\em dynamic} saddle point,
characterized by a time-dependent propagator of the bosonic field $B_i$
conjugate to $f^{\dagger}_{\alpha}c_{\alpha i} (0)$.
One finds the following integral equations for the fermionic and
bosonic Green's functions
$G_f(\tau) = -\langle {\rm T}_\tau f_\alpha(\tau) f_\alpha^\dagger(0) \rangle$,
$G_B(\tau) =  \langle {\rm T}_\tau B_i(\tau) B_i^\dagger(0) \rangle $:
\begin{equation}
\label{sp}
\Sigma_f(\tau)=\gamma G_0^{(\rm f)}(\tau) G_B(\tau) ,\,\,\,
\Sigma_B(\tau)= -G_0^{(\rm f)}(-\tau) G_f(\tau) \,,
\end{equation}
where $G_0^{(\rm f)}$ is the local Green's function of the fermionic bath,
and the self-energies $\Sigma_f$ and $\Sigma_B$ are defined by:
\begin{eqnarray}
\label{defsigmaf}
G_f^{-1}(i\omega_n) &=& i\omega_n+\lam-\Sigma_f(i\omega_n)  \\
\label{defsigmab}
G_B^{-1}(i\nu_n) &=& {{1}\over{\Jb}}-\Sigma_B(i\nu_n) \,.
\end{eqnarray}
In these expressions $\omega_n=(2n+1)\pi/\beta$ and $\nu_n=2n\pi/\beta$
denote fermionic and bosonic Matsubara frequencies.
The third saddle point equation fixes the impurity ``occupation'' and thus
determines $\lam$:
\begin{equation}
\label{eqlambda}
G_f(\tau=0^-) = q_0 \,.
\end{equation}
Eqs.~(\ref{sp},\ref{defsigmaf},\ref{defsigmab}) are identical in structure
to the usual NCA equations \cite{CR}, but the physics is changed by the constraint
eq.~(\ref{eqlambda}) which keeps track of the choice of the impurity
spin representation.

The analysis of the NCA equations for the pseudogap case has been presented
in Ref.~\onlinecite{OSPG}.
As expected for a multi-channel Kondo model the exactly screened phase at large
$\Jb$ is replaced by an overscreened one.
For $r>0$ this overscreened phase is bounded by
second-order transitions to phases with either a free spin moment (LM, at small $\Jb$) or a free
channel moment (the latter only exists for p-h asymmetry and for large $\Jb$).
In both the overscreened phase and at criticality the auxiliary-particle propagators
$G_f$ and $G_B$ follow power laws at low energy.
The phase transitions display $r$-dependent exponents -- naturally these differ from the
ones of the SU(2) single-channel model.
Various critical properties are reproduced correctly, e.g., the $\omega^{-r}$
divergence of the conduction electron $T$ matrix at criticality, see below.
In the p-h asymmetric case $r=1$ is recovered as upper-critical
dimension.

\begin{figure}
\epsfxsize=3.1in
\centerline{\epsffile{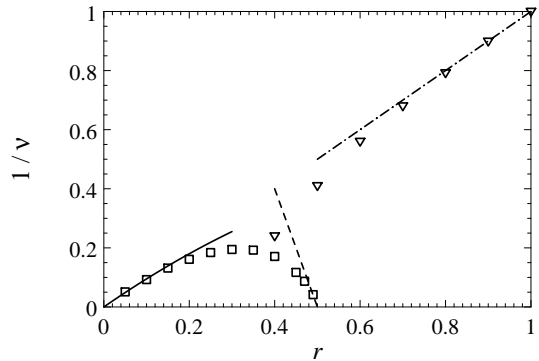}}
\caption{
Inverse correlation length exponent $1/\nu$ obtained from NRG,
at both the symmetric (squares) and asymmetric (triangles) critical points,
together with the analytical RG results from the expansions
in $r$ (solid), in $(\frac{1}{2} -r)$ (dashed), and
in $(1-r)$ (dash-dot), eq. (\ref{nu_pg}).
After Ref.~\protect\onlinecite{MVLF}.
}
\label{fig:nuz}
\end{figure}

\subsection{Results for observables}
\label{sec:pgobs}

Let us quote some important observables near criticality for the SU(2)-symmetric pseudogap
Anderson and Kondo models -- these results have been obtained using NRG and by the
three $\epsilon$-type expansions together with renormalized perturbation
theory.

The correlation length exponent $\nu$ (\ref{defnu}) can be obtained from
the perturbative RG by expanding the beta functions in the vicinity of the
fixed point.
The results are
\begin{equation}
\frac{1}{\nu} = \left\{
\begin{array}{ll}
r - \frac{r^2}{2} +  {\cal O}(r^3)              & \mbox{SCR},~r\ll 1 \\[3mm]
2-4r + {\cal O}\!\left([\frac 1 2 -r]^2\right)  & \mbox{SCR},~\frac 1 2 - r \ll 1 \\[3mm]
r + {\cal O}\!\left([1-r]^2\right)              & \mbox{ACR},~1-r\ll 1
\end{array}
\right. .
\label{nu_pg}
\end{equation}
At the symmetric critical fixed point $\nu$ diverges for both $r\to 0^+$
and $r\to 1/2^-$.
For $r\geq 1$ the transition is a level crossing, formally $\nu=1$.
Fig. \ref{fig:nuz} shows a comparison of these results for $\nu$
with NRG results.
Other exponents can also be evaluated perturbatively, see Ref.~\onlinecite{MVLF},
and the full set of exponents (\ref{exponents}) can be derived from hyperscaling
relations which are valid for $r<1$.

\begin{figure}
\epsfxsize=3.1in
\centerline{\epsffile{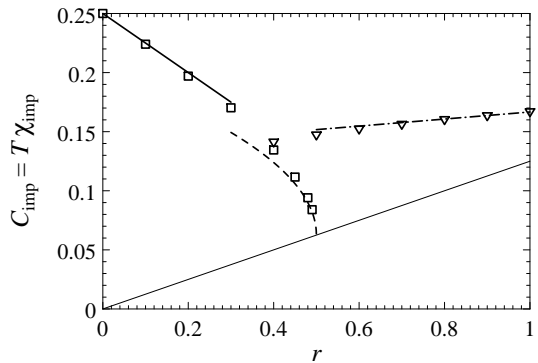}}
\caption{
Numerical data for the impurity susceptibility, $T\chi_{\rm imp}$,
at both the symmetric (squares) and asymmetric (triangles) critical points,
together with the renormalized perturbation theory results from the expansions
in $r$ [solid], in $(\frac{1}{2} -r)$ [dashed], and in $(1-r)$ [dash-dot].
After Ref.~\protect\onlinecite{MVLF}.
}
\label{fig:tchi}
\end{figure}

In the quantum critical regime unconventional behavior corresponding to a fractional moment
can be observed. As explained in Sec.~\ref{sec:obs} the impurity entropy $S_{\rm imp}$ and
the impurity Curie moment $T\chi_{\rm imp}$ (Fig.~\ref{fig:tchi})
are universal functions of $r$ in the low-temperature limit \cite{GBI,MVLF}.

Let us briefly mention our result for the conduction electron $T$ matrix,
describing the scattering of the $c$ electrons off the impurity.
At criticality it follows a power law (\ref{tmatpower}),
and one finds the {\em exact} result $\eta_T=1-r$ for $r<1$,
i.e., for all interacting fixed points
considered above the $T$ matrix follows $T(\omega)\propto\omega^{-r}$.
This perfectly agrees with NRG calculations \cite{bullapg}.
Furthermore, at the upper-critical dimension
we find ${\rm Im}\,T(\omega)\propto 1/(\omega |\log{\omega}|^{2})$ --
this applies, e.g., to a Kondo impurity in a $d$-wave superconductor.

\subsection{Applications}

The Kondo effect in a non-metallic host is most conveniently studied
using superconductors.
As above, both impurity spins embedded in a bulk material and interacting
quantum dots are candidate realizations.
A recent experiment \cite{basel} employed a carbon nanotube dot coupled
to Nb leads, which can be switched between $s$-wave superconducting
and normal states by a small magnetic field.
A sharp crossover in the transport properties was found as a function
of $T_K/\Delta$, where $T_K$ is the normal-state Kondo temperature of the dot
and $\Delta$ the gap of the superconductor, consistent with the expected first-order
transition within a hard-gap Kondo model.
Quantum dots coupled to unconventional superconductors remain to be realized.
Note that a complete description of the transport requires to account for
multiple Andreev reflections.
Other suggestions for pseudogap Kondo physics include quantum dots coupled
to either interacting one-dimensional electron liquids \cite{kim}
or to disordered mesoscopic conductors \cite{hopkinson}.

Impurity moments embedded in bulk unconventional superconductors
(or other systems with nodal order parameters)
represent another realization of pseudogap Kondo physics.
Indeed, in $d$-wave high-$T_c$ superconductors ($r=1$ here)
non-trivial Kondo-like behavior has been observed associated with the magnetic
moments induced by non-magnetic Zn or Li impurities~\cite{bobroff}.
In underdoped superconducting samples the local susceptibility as measured
by NMR on the impurity site displays a Curie law which turns into a Curie-Weiss
law at optimal doping.
This apparent transition from an unscreened to a screened moment is likely
driven by a change of either the size of the superconducting gap, i.e.,
a change of the dimensionless coupling in the pseudogap Kondo model,
or of increasing strength of the bulk spin fluctuations upon underdoping
which tend to suppress Kondo screening (see Sec.~\ref{sec:bfk}).
STM experiments \cite{seamus} show a large peak at small bias in the local density
of states near the impurity site -- this has been interpreted as arising from the Kondo
screening of the impurity-induced moment, and explicit calculations within a p-h
asymmetric pseudogap Kondo model have been presented \cite{tolya,MVRB}.


\section{Spin-boson model}
\label{sec:spb}

After having described fermionic impurity models with intermediate-coupling fixed points,
let us turn to systems of quantum impurities coupled to {\em bosonic} baths,
which are an equally interesting model class.
They have been first introduced in the context of the description of dissipative
dynamics in quantum systems \cite{leggett}.
The simplest realization is the so-called spin-boson model, describing a
two-level system coupled to a single bath of harmonic oscillators:
\begin{equation}
{\cal H}_{\rm SB}
=-\frac{\Db}{2}\sigma_{x}+\frac{\epsilon}{2}\sigma_{z}+
\sum_{i} \omega_{i}
     a_{i}^{\dagger} a_{i}
+ \frac{\sigma_{z}}{2} \sum_{i}
    \lambda_{i}( a_i + a_i^{\dagger} ) \,.
\label{eq:sbm}
\end{equation}
It describes a spin, tunneling between
$|\uparrow\rangle$ and $|\downarrow\rangle$ via $\Db$, and being damped
by the coupling to the oscillator bath.
The coupling between spin and bath is completely specified by the bath
spectral function
\begin{equation}
    J(\omega)=\pi \sum_{i}
\lambda_{i}^{2} \delta\left( \omega -\omega_{i} \right) \,.
\end{equation}
Of particular interest are power-law spectra
\begin{equation}
  J(\omega) = 2\pi \alpha \omega_c^{1-s} \omega^s\,,~ 0<\omega<\omega_c\,,\ \ \ s>-1
\label{power}
\end{equation}
where $\omega_c$ is a cutoff, and the dimensionless parameter
$\alpha$ characterizes the coupling or dissipation strength.

The case of $s\!=\!1$ corresponds to the well-studied
ohmic spin-boson model \cite{leggett}, which shows a Kosterlitz-Thouless
quantum transition, separating a localized phase at $\alpha \geq \alpha_c$
from a delocalized phase at $\alpha<\alpha_c$ \cite{leggett,weiss}.
In the localized regime, the tunnel splitting between the two
levels renormalizes to zero, i.e., the system gets trapped in one
of the states $|\uparrow\rangle$ or $|\downarrow\rangle$,
whereas the tunnel splitting stays finite in the delocalized
phase.
In the limit $\Db \ll \omega_c$ the transition occurs at $\alpha_c=1$.
The super-ohmic situation, $s>1$ is known to show weakly damped dynamics for any
dissipation strength $\alpha$, i.e., the two-level system is always delocalized in the
zero-temperature limit, provided that $\Db\neq 0$.
Much less is known about the sub-ohmic case ($s<1$) -- this turns out to be very interesting
and will be discussed below.

\subsection{Spin-boson model: Phase diagram and NRG}

In the following we focus on the sub-ohmic exponent range, $0\leq s< 1$.
It has recently been established that a {\em continuous} quantum transition
between a localized and a delocalized phase exists for all bath exponents
$0<s<1$ \cite{bosnrg}.
We emphasize that this transition does not appear in the popular non-interacting
blip (NIBA) approximation \cite{weiss}.

\begin{figure}[!t]
\epsfxsize=3.2in
\centerline{\epsffile{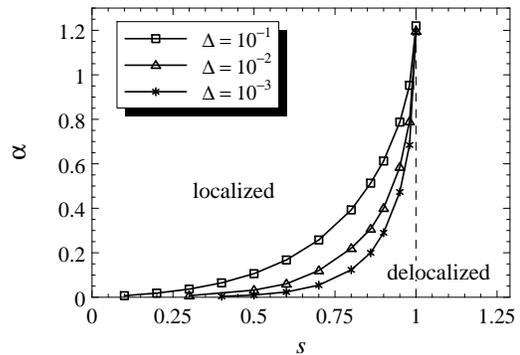}}
\caption{
Phase diagram for the transition between delocalized
($\alpha < \alpha_c$) and localized phases ($\alpha > \alpha_c$)
of the spin-boson model (\protect\ref{eq:sbm})
for bias $\epsilon=0$ and various values of $\Db$,
deduced from NRG.
After Ref.~\protect\onlinecite{bosnrg}.
}
\label{fig:spb_phd1}
\end{figure}

The two stable phases of the sub-ohmic spin-boson model are conventional:
The delocalized phase has a unique ground state and a finite susceptibility $\chi^{zz}$,
i.e., the level splitting is finite in the low-temperature limit.
In contrast, the localized phase has a residual entropy $S_{\rm imp} = \ln 2$
and $\chi^{zz}$ follows a Curie law of the form (\ref{mloc}).
For fixed $s$ there exists a single critical fixed point with $s$-dependent exponents;
as function of $s$ these fixed points form a line which terminates in the Kosterlitz-Thouless
transition point at $s\!=\!1$.
The transition also disappears in the limit $s\to 0^+$, and the model is
always in a localized ground state for $s\leq0$.
Numerical results for the phase diagram are shown in Fig.~\ref{fig:spb_phd1}.

It is generally believed that the spin-boson model is equivalent to
a one-dimensional classical Ising model with long-range interactions \cite{leggett,emery,koster},
specifically the bath with $\omega^s$ damping rate corresponds to an
Ising interaction falling off as $1/r^{1+s}$.
This quantum--classical mapping will be discussed in more detail in
Sec.~\ref{sec:qucl}:
while this equivalence holds for $s>1/2$ including the ohmic case,
it has recently been shown \cite{quclfail} that it is violated for $s<1/2$.

\subsection{Spin-boson model: RG near $s=1$}

We now turn to analytical approaches to describe the spin-boson phase transition.
Power counting about the $\Db=\alpha=0$ fixed point
shows that the coupling $\alpha$ is marginal at $s=1$,
suggesting that a perturbative RG may be possible.
However, for the spin-boson model, i.e., Ising symmetry, one encounters
a flow of $\alpha$ towards strong coupling for $s<1$.
Progress can be made using a different formulation of the partition function,
namely using a kink gas representation,
where the kinks represent spin flips along the imaginary time axis
or equivalently Ising domain walls \cite{koster}.
A perturbative RG analysis of this Ising model has been performed by
Kosterlitz \cite{koster}.
This expansion, controlled by the smallness of the kink fugacity, is done around
the ordered phase of the Ising model, corresponding to the localized fixed point
of the spin-boson quantum problem.
Carrying over these results to the spin-boson model, ones arrives at RG equations:
\begin{eqnarray}
\beta(\alpha) &=& \alpha ( \Delta^2 + s - 1 ) \,,\nonumber \\
\beta(\Delta) &=& -\Delta ( 1 - \alpha ) \,,
\label{eq:rg}
\end{eqnarray}
valid for small $\Delta$,
where $\Delta = \Db/\omega_c$ is the dimensionless tunneling strength.
The RG flow is sketched in Fig. 1 of Ref.~\onlinecite{koster}.
For $s=1$, these equations are equivalent to the ones known from
the anisotropic Kondo model, and describe a Kosterlitz-Thouless transition,
with a fixed line $\Delta = 0$, $\alpha\geq 1$.
For $s<1$ there is an unstable fixed point at $\alpha=1$, $\Delta^2 = 1-s$;
clearly it is perturbatively accessible for small values of
$(1\!-\!s)$ only.

\subsection{Spin-boson model: RG near $s=0$}
\label{sec:rgs0}

A different RG expansion can be performed around
the delocalized fixed point of the spin-boson model,
corresponding to finite $\Db$ and infinitesimal $\alpha$.
For convenience we shall assume equal couplings, $\lambda_i \equiv \lambda$,
then the energy dependence of $J(\omega)$ is contained in the density of states of
the oscillator modes $\omega_i$, and we have $\alpha \propto \lambda^2$.
Eigenstates of the impurity are $|\!\rightarrow_x\rangle$ and $|\!\leftarrow_x\rangle$,
with an energy splitting of $\Db$.
The low-energy Hilbert space contains the state $|\!\rightarrow_x\rangle$
only, and interaction processes with the bath arise in second-order
perturbation theory, proportional to $\kappa_0 = \lambda^2/\Db$.
Power counting shows that $\kappa_0$ is marginal
at $s=0$.
Performing RG within the low-energy sector one finds the RG beta function
for the renormalized second-order tunneling amplitude:
\begin{eqnarray}
\beta(\kappa) = s \kappa - \kappa^2 \,.
\end{eqnarray}
Besides the stable delocalized fixed point $\kappa=0$ this
flow equation displays an infrared unstable fixed point at
$\kappa^\ast = s$ which controls the transition between the delocalized and
localized phases.
It is worth pointing out that no (dangerously) irrelevant variables are present
in this theory.

As an aside, we note that at $s=0$ the bath coupling is marginally relevant.
Therefore the impurity is always localized as $T\to 0$, with a localization
temperature given by $T^\ast = \omega_c \exp(-\Db\omega_c / \lambda^2)$.

\subsection{Results for observables}

We now turn to results for various critical properties of the
spin-boson model, determined using the two RG expansions and via
bosonic NRG.

\begin{figure}[t]
\epsfxsize=3.1in
\centerline{\epsffile{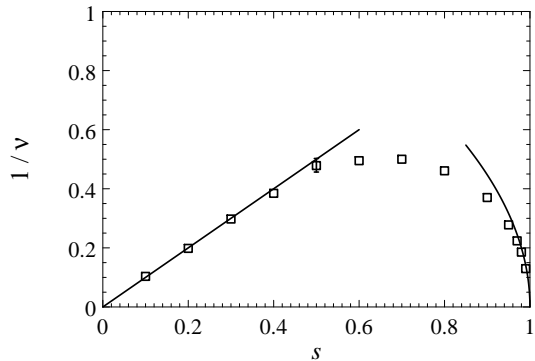}}
\caption{
Numerical data \cite{quclfail} for the correlation length exponent $1/\nu$ obtained from NRG,
together with the two perturbative RG results in Eq. (\ref{nuz_s})
which are asymptotically valid near $s=1$ and $s=0$, respectively.
}
\label{fig:nu_spb}
\end{figure}

The correlation length exponent $\nu$ (\ref{defnu}) is obtained by
expanding the RG beta functions near the critical fixed point.
One finds:
\begin{equation}
1/ \nu = \left\{
\begin{array}{ll}
\sqrt{2(1-s)} + {\cal O}(1-s)   & ~1-s\ll 1 \\[3mm]
s + {\cal O}(s^2)               & s\ll 1
\end{array}
\right. .
\label{nuz_s}
\end{equation}
These results in excellent agreement with NRG data, as shown in Fig.~\ref{fig:nu_spb}.
Interestingly, the correlation length exponent diverges for both $s\to 0^+$ and $s\to 1^-$,
where the second-order transition disappears.
Thus the transition is bounded by {\em two lower-critical dimensions},
somewhat similar to the p-h symmetric pseudogap Kondo model
described in Sec.~\ref{sec:pgk}.
However, for $s<1$ and $r>0$ the two problems are in different universality
classes \cite{MVLF}, as can be seen e.g. by comparing the correlation length exponents
in Figs.~\ref{fig:nuz} and \ref{fig:nu_spb}.

For small $\alpha$, $\Db$, Eq. (\ref{eq:rg}) yields for the phase boundary
\begin{equation}
\alpha_c \propto \Db^{1-s} ~~ \mbox{for}~ \Db\ll\omega_c
\label{alpcrg}
\end{equation}
valid for all $0<s<1$, which compares well with NRG \cite{bosnrg}.
We also quote a few results for the critical exponents associated with
a local field (\ref{exponents}), obtained by the small-$s$ expansion
of Sec.~\ref{sec:rgs0}:
\begin{eqnarray}
\gamma &=& 1 + {\cal O}(s) , \nonumber\\
1/\delta &=& 1 - 2 s + {\cal O}(s^2) \,.
\label{delta}
\end{eqnarray}
Further, one finds the {\em exact} result $x = y = s$ which relies on the
bath bosons being non-interacting \cite{quclfail}.
Thus the local spin correlation function $C(\omega)$ diverges as
$\omega^{-s}$ at criticality, whereas it follows $\omega^{s}$
in the disordered phase.

\begin{figure}[t]
\epsfxsize=3.1in
\centerline{\epsffile{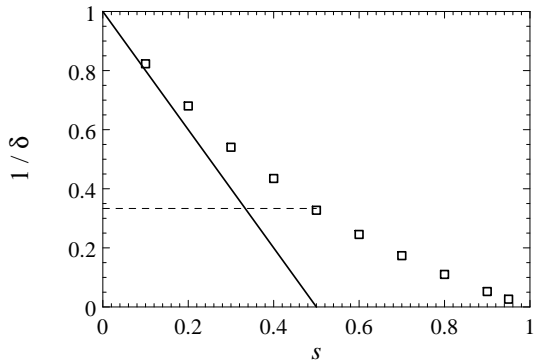}}
\caption{
Numerical data for the critical exponent $1/\delta$ (\ref{exponents})
obtained from NRG, together with the RG result (\ref{delta}).
The dashed shows the mean-field value $\delta = 3$ of the one-dimensional
Ising model with $1/r^{1+s}$ long-range interaction.
After Ref.~\protect\onlinecite{quclfail}.
}
\label{fig:delta_spb}
\end{figure}

We note that all critical exponents obey hyperscaling which has been verified by NRG;
we conclude that the critical fixed point is interacting for all $0<s<1$.
The exponents $\beta$ and $\delta$ (Fig.~\ref{fig:delta_spb}) are distinct from the
ones of the long-range Ising model with $1/r^{1+s}$ interaction for $s<1/2$ which
displays mean-field behavior there. The obvious failure of the quantum--classical
mapping will be discussed in Sec.~\ref{sec:qucl}.

\subsection{Applications}

Spin-boson and dissipative impurity models \cite{leggett,weiss}
have applications in many fields like glass physics, mechanic friction,
damping in electric circuits, electron transfer in biological molecules etc.
Re-newed interest in spin-boson models arises in the quantum computation,
for modelling the coupling of qubits to a noisy environment and the
associated decoherence processes.
Interestingly, the description of $1/f$ noise in
electrical cicuits leads to sub-ohmic damping with $s=0$ (at least over
a certain range of energies).

Modifications of standard spin-boson physics include the influence
of localized modes which interact with two-level systems -- those modes
can be represented by a discrete spin system, leading to so-called central
spin models.
Electron transport through quantum dots under the influence of external
noise or degrees of freedom like phonons involves the combined effect
of tunneling from electronic leads and bosonic damping,
with competing interaction channels -- this naturally leads
to Bose-Fermi Kondo models described in Sec.~\ref{sec:bfk}.


\section{Bose Kondo model}
\label{sec:bk}

A class of models which is closely related to the spin-boson model,
termed Bose Kondo models, describe magnetic impurities
coupled to spinful bosons \cite{subirbk}.
Typically these bosons repesent collective bulk spin fluctuations.
The Hamiltonian is conveniently written as
${\cal H} = {\cal H}_{\rm BK} + {\cal H}_{\rm b}^{(\rm b)}$, with
\begin{equation}
{\cal H}_{\rm BK} =
\gamma_0 \SSS_\alpha \phi_\alpha(0) \,.
\label{eq:bk}
\end{equation}
In the simplest case, the bosonic bath consists of free vector bosons,
\begin{equation}
{\cal H}_{\rm b}^{(\rm b)} = \sum_{q\alpha} \omega_q a_{q\alpha}^\dagger a_{q\alpha}
\end{equation}
describing collective spin fluctuations in the $d$-dimensional host material, with
a dispersion $\omega_q^2 = m^2 + c^2 q^2$.
At zero temperature, the mass $m$ coincides with the bulk spin gap $\Delta_s$.
A gap $\Delta_s>0$ describes a quantum paramagnet, $\Delta_s\!=\!0$ corresponds to a bulk quantum
critical point between the paramagnet and an antiferromagnetically ordered phase;
the momentum $q$ is measured relative to the ordering wavevector $Q$.
The field $\phi_\alpha(0)$ in Eq. (\ref{eq:bk}) represents the local orientation of the
antiferromagnetic order parameter, it is given by
$\phi_\alpha(0) = \sum_q (a_{q\alpha} + a_{-q\alpha}^\dagger) /
\sqrt{\omega_q/J}$, where $J$ is the bulk exchange constant, i.e.,
in the parametrization of Eq. (\ref{eq:sbm}) we have
$\lambda_q \propto \omega_q^{-1/2}$.
Importantly, in this model the SU(2) symmetry is preserved, contrary to the
spin-boson model (\ref{eq:sbm}).

In a paramagnetic phase of the bulk, the $a_q$ are gapped, and the
low-energy properties of ${\cal H}_{\rm BK}$ are that of a free spin.
In contrast, at a magnetic bulk critical point in $d$ dimensions
the $a_q$ obey a gapless spectrum of the form (\ref{power}), with
exponent $s=d-2$.
Note that in dimensions $d<3$, boson self-interactions are relevant in the RG sense
at the critical point, and then a $\phi^4$ theory formulation
becomes more appropriate \cite{science,vbs}:
\begin{equation}
{\cal H}_{\rm b}^{(\rm b)} = \int d^d x \left[ \frac{ \pi_{\alpha}^2 + c^2
(\nabla \phi_{\alpha})^2 + s\phi_{\alpha}^2}{2} + \frac{u_0}{4!}
(\phi_{\alpha}^2)^2 \right]
\end{equation}
where $s$ is the tuning parameter, $c$ a velocite,
$u_0$ the bulk non-linear interaction, and
$\pi_{\alpha}$ the momentum conjugate to $\phi_\alpha$, hence
\begin{equation}
[\phi_{\alpha} (x,t), \pi_{\beta} (x',t)] = i \delta_{\alpha
\beta} \delta^d (x-x') \,.
\end{equation}
In a magnetically ordered bulk phase, the Hamiltonian includes
the local exchange field from the static order parameter, and
only transverse magnetic fluctuations remain gapless.

In the representation (\ref{eq:bk}) of the Bose Kondo problem the complexity of
the impurity lies in the spin commutation relations of $S_\alpha$;
the model can also be written using a spin coherent state path integral,
then the only impurity non-linearity appears in the Berry phase term \cite{vbs,subirbk}.

\subsection{Bose Kondo model: RG near $s=1$}

The model (\ref{eq:bk}) can be analyzed using a perturbative RG
expansion in $\gamma_0$ and the bulk interaction $u_0$, which are
both marginal at $s=1$ ($d=3$).
It is important to treat both couplings on an equal footing as there is a
non-trivial ``interference'' between $u_0$ and $\gamma_0$, which appears at two-loop order,
and the interaction $u_0$ significantly modifies the magnetic environment coupling to the
impurity.
In the magnetic context it is {\em not} permissible to treat the environment as a
Gaussian quantum noise, and focus only on its Kondo-like coupling
to the impurity.

A systematic RG analysis of ${\cal H}_{\rm BK}$ was carried out \cite{science,vbs},
resulting in the following RG equation for the renormalized impurity
coupling $\gamma$:
\begin{equation}
\beta(\gamma) = -\frac{1-s}{2} \gamma + \gamma^3 - \gamma^5
+ \frac{5 u^2 \gamma}{144} + \frac{u \gamma^3 \pi^2}{3} [ S(S+1)-1/3]
\label{eps21}
\end{equation}
to two-loop order, and $u$ is the renormalized bulk interaction.
For $s<1$, i.e., $d<3$, there is a {\em stable} intermediate coupling fixed point
at ${\gamma^\ast}^2 = (1-s)/2 + {\cal O}([1-s]^2)$ --
this ``bosonic fluctuating'' fixed point (B-FL)
displays universal local-moment fluctuations as detailed below.
Note that this is in stark contrast to the model with Ising
symmetry which flows to strong coupling in the same parameter regime --
this strong-coupling fixed point is apparently prohibited here by the
competition between the multiple baths.
The B-FL fixed point is unstable w.r.t. SU(2) symmetry breaking,
but the existence of an intermediate-coupling fixed point is preserved for XY symmetry.

For $s\geq 1$ the coupling $\gamma$ is (marginally) irrelevant.
Here, the low-temperature properties of ${\cal H}_{\rm BK}$ are equivalent
to an asymptotically free spin.

\subsection{Bose Kondo model: RG near $s=-1$}

An alternative approach to the dynamics of an impurity embedded in a
bulk magnet is provided by a representation of the bulk spin fluctuations
using a non-linear sigma model.
The fluctuations of the order parameter amplitude, $\phi_{\alpha}^2$,
become irrelevant in low dimensions,
and a hard-spin representation by a unit-length field $N_{\alpha} (x, \tau)$,
capturing only angular fluctuations of the Neel order parameter, is appropriate.
The non-linear sigma model formulation allows for an RG expansion
of the bulk quantum critical properties in $(1\!+\!\epsilon)$ dimensions.

The Bose Kondo problem can be tackled using such a
fixed-length representation \cite{sv}.
Interestingly, in the scaling limit the quantum impurity behaves as
if it is in the $\gamma \rightarrow \infty$ limit, and hence the universal
nature of the coupling between the bulk and impurity is explicit.
The properties of the stable impurity fixed point at bulk criticality
can be computed in a systematic expansion in $(d-1)$
(i.e., near $s=-1$ in terms of the bath spectrum exponent).
All results were found to be consistent with those of the $(3-d)$
expansion of the previous subsection.

\subsection{Bose Kondo model: large-$N$}
\label{sec:bklargen}

Generalizing the impurity spin to SU($N$) symmetry allows for
a large-$N$ limit corresponding to a dynamic saddle point \cite{vbs}.
With the impurity representation (\ref{imprep})
the Hamiltonian for the impurity takes the form
\begin{eqnarray}
\mathcal{H}_{\rm BK} &=&  \sum_{q\alpha\beta}
      \lambda_q
      f_\alpha^\dagger f_\beta
      (b_{q\alpha\beta} + b_{-q\alpha\beta}^\dagger)
+ \lam f^{\dagger}_{\alpha} f_\alpha
\label{himpN}
\:.
\end{eqnarray}
For $N\!\to\!\infty$ we obtain an integral equation for the
$f$ fermion Green's function:
\begin{equation}
\Sigma_f(\tau)= G_0^{(\rm b)}(\tau) G_{f}(\tau)
\label{ncasigma}
\end{equation}
where $G_0^{(\rm b)}$ is the local Green's function of the bath,
$G_0^{(\rm b)}(\tau) = -\sum_q \lambda_q^2 \langle {\rm T}_\tau (b_q+b_{-q}^\dagger)(\tau) (b_q^\dagger+b_{-q})(0)
\rangle$,
and the self-energy $\Sigma_f$ is defined by:
\begin{equation}
\label{defsigma}
G_f^{-1}(i\omega_n) = i\omega_n - \lam - \Sigma_f(i\omega_n)
\:.
\end{equation}
These equations correspond to the summation of all self-energy diagrams
with non-crossing bath boson lines, and can be understood as the
bosonic analogue of the fermionic NCA equations (\ref{sp},\ref{defsigmaf},\ref{defsigmab}).

The equations (\ref{ncasigma},\ref{defsigma}) have been analyzed in Ref.~\onlinecite{vbs}.
One obtains a $G_f$ solution with a power-law behavior at low energies,
corresponding to an intermediate-coupling fixed point.
Many of the properties are similar to the SU(2) case with non-interacting bosons,
analyzed above using RG.
Interestingly, the NCA-type approach can be extended to a finite concentration
of impurities and allows to capture the feedback of the impurity
scattering to the bulk spin excitations \cite{vbs}.

\subsection{Results for observables}
\label{sec:bkobs}

Although the Bose Kondo model does not display a quantum phase transition,
we quote a few results for the intermediate-coupling fixed point obtained
from the RG expansion. Some of the properties for interacting bosons in $d=2$
have been also investigated by quantum Monte Carlo simulations of two-dimensional
lattice antiferromagnets \cite{troyer,sandvik}.

The structure of the fixed point implies that the impurity shows
universal local-moment fluctuations:
the local susceptibility $\chi_{\rm loc}$ obeys a scaling form (\ref{scalchi1})
with an anomalous exponent $\eta_\chi > 0$ for $s<1$,
which means that local spin correlations at $T=0$ are characterized by a
power law \cite{bfk,bfknew,science},
$\langle{\hat S}(\tau){\hat S}\rangle \propto \tau^{-\eta_\chi}$.
For non-interacting bath bosons, one obtains the {\em exact} result
$\eta_\chi = 1-s$ [or $x=y=s$ according to (\ref{exponents})].
Interestingly, this exact exponent implies that the impurity
fluctuates faster than its environment for $s<0$,
which violates a theorem due to Griffiths, stating that the impurity
has to fluctuate slower than the environment, $\eta_\chi < 1+s$.
Thus, the Bose Kondo problem with non-interacting bosons likely
changes qualitatively at $s=0$, however, details are unknown.
In contrast, higher powers of $(1-s)$ appear in the perturbative
expansion of $\eta_\chi$ for interacting bosons, and the results
are expected to fulfill the above inequality for all $-1<s<1$.

Other observables at the intermediate-coupling fixed point can be
associated with the properties of a fractional spin:
The Curie contribution to the impurity susceptibility is
$T\chi_{\rm imp}= {\cal C}_{\rm imp}$,
and the impurity entropy $S_{\rm imp}$ is a finite constant and larger than $\ln 2$.
Both ${\cal C}_{\rm imp}$ and $S_{\rm imp}$ are universal functions
of $s$ (or the dimensionality $d$ of the bath), but depend on
whether the bosons are interacting and non-interacting.
Related theoretical results were
obtained recently \cite{castro} in magnetically ordered states in
the presence of spin anisotropy, and in Ref.~\onlinecite{sushkov2}.

We summarize that the properties of the Bose Kondo problem with a bath given by a
quantum critical magnet, i.e., with interacting bosons, appear to
vary smoothly for $1<d<3$ (i.e., $-1<s<1$).
For $d=1$ the problem is ill-defined as the bulk transition disappears
(and the environment at the putative transition ceases to fluctuate),
and for $d\geq 3$ the impurity coupling is (marginally) irrelevant,
resulting in an asymptotically free spin.
In contrast, in the Bose Kondo model with a non-interacting
(Gaussian) bath the impurity behavior is believed to change
qualitatively at $s=0$, i.e., non-trivial intermediate coupling physics
is only present for $0<s<1$.

\subsection{Applications}

Vector baths naturally appear in the context of bulk spin fluctuations.
Thus impurity moments embedded in magnets are naturally described
by Bose Kondo models -- such models become appropriate
when the environment of the impurity is in the vicinity of a magnetic ordering
transition, and there are low-energy spin excitations in the bulk;
the latter may be viewed as excitonic particle-hole bound states
of a metal/insulator/superconductor which peel off below the
continuum of a pair of fermionic particles or holes.
As discussed in Refs.~\onlinecite{science,vbs} the self-interaction of the bulk bosons
cannot be neglected below the upper-critical dimension of the bulk, as
it also changes the impurity physics.
Concrete realizations are magnetic atoms in insulating bulk magnets.
At low energies, these considerations can be extended to $d$-wave
superconductors, and can describe non-magnetic impurities like Zn or Li
in cuprates, which are known to induce magnetic moments.
For a finite impurity concentration,
the universal interaction between impurity moments and host spin fluctuations
leads to universal impurity damping of spin fluctuations in cuprate
superconductors -- corresponding signatures have been observed in inelastic
neutron scattering in Zn-doped YBa$_2$Cu$_3$O$_7$ \cite{vbs}.

Ref.~\onlinecite{soc} discussed the photoemission spectrum of a conduction
electron in an insulator in the vicinity of the magnetic
transition \cite{assaad}.
Let us focus on the minimum energy of its dispersion.
While away from the critical point a sharp quasiparticle pole appears,
the interaction with the magnetic modes leads to a universal
damping of this pole at the magnetic bulk transition,
resulting in a power-law spectrum
$G(\omega) \sim 1/(\omega - \varepsilon_0 )^{1 - \eta_f}$.
This behavior reflects an orthogonality catastrophy in the spin
sector, and is the bosonic analogue of the familiar X-ray edge problem
in a Fermi liquid \cite{soc}.

Spins coupled to bulk spin fluctuations can also occur in
quantum dots coupled to leads with magnetic collective modes;
taking into account the fermionic degrees of freedom as well leads
to Bose-Fermi Kondo models as described in Sec.~\ref{sec:bfk}.


\section{Bose-Fermi Kondo model}
\label{sec:bfk}

Novel phenomena occur for magnetic impurities coupled
to {\em both} a fermionic and a bosonic bath, where the bosons represent
e.g. collective host spin excitations.
In the resulting Bose-Fermi Kondo model the two interactions compete in a
non-trivial manner~\cite{bfk,bfknew}, and fermionic Kondo
screening can be strongly suppressed by host spin fluctuations.
The Hamiltonian is a straightforward combination of the
bosonic and fermionic Kondo models,
${\cal H} = {\cal H}_{\rm BFK} + {\cal H}_{\rm b}^{(\rm b)} + {\cal H}_{\rm b}^{(\rm f)}$,
with an impurity spin $S=1/2$ and
\begin{eqnarray}
{\cal H}_{\rm BFK} =
\gamma_0 \SSS_\alpha \phi_\alpha(0)
+ \Jb \SSS_\alpha s_\alpha(0)
\label{eq:bfk}
\end{eqnarray}
with notations as above.
The spin-1 order parameter field $\phi_\alpha$ and the spin-1/2 fermions $c_{k\sigma}$,
with local spin density $s_\alpha(0)$,
are described by the baths ${\cal H}_{\rm b}^{(\rm b)}$ and ${\cal H}_{\rm b}^{(\rm f)}$,
respectively.
Most interesting is again the case of a bosonic bath with zero or small gap,
corresponding to the vicinity to a magnetic quantum critical point in the
bulk.

\subsection{Bose-Fermi Kondo model: Phase diagram and RG}

We now discuss the phase diagram of the Bose-Fermi Kondo model,
based on perturbative RG and large-$N$ results.
We restrict ourselves here to a metallic density of states for the
fermions (the pseudogap case was discussed in Ref.~\onlinecite{MVMK}).

\begin{figure}[!t]
\epsfxsize=2.7in
\centerline{\epsffile{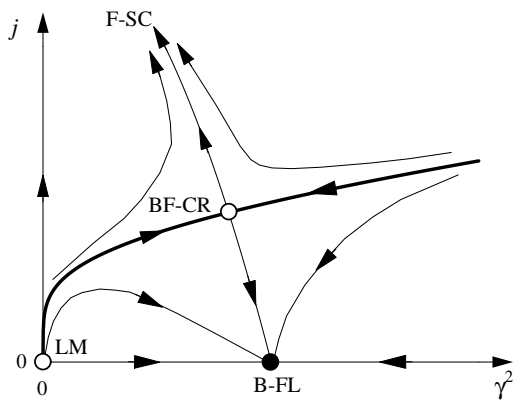}}
\caption{
RG flow diagram for the Bose-Fermi Kondo model.
The two axes denote the renormalized couplings to the fermionic ($j$)
and bosonic ($\gamma$) baths, respectively.
LM is the local-moment fixed point of a decoupled impurity,
F-SC the fermionic Kondo-screened fixed point, B-FL the intermediate-coupling
fixed point described in Sec.~\protect\ref{sec:bk}, and FB-CR is critical
fixed point of the Bose-Fermi Kondo problem.
A metallic density of states of the fermionic bath is assumed,
and the bosonic bath has $s<1$, i.e., $d<3$ in the magnetic context.
After Ref.~\protect\onlinecite{bfknew}.
}
\label{fig:bfkflow}
\end{figure}

For a gapless bosonic bath with spectrum $\omega^s$, $s<1$, corresponding
to a magnetic critical point in $2+s$ dimensions, both the fermionic
and bosonic couplings are relevant, and compete.
For large $\Jb$ fermionic Kondo screening wins, resulting in a fully screened
spin. Large $\gamma_0$ can completely suppress Kondo screening,
driving the system into the intermediate-coupling fixed point of the
Bose-Kondo problem, with universal local-moment fluctuations (Sec.~\ref{sec:bk}.
The competition is captured by the RG equations \cite{bfk,bfknew,MVMK}:
\begin{eqnarray}
\beta(\gamma) &=& -\frac{(1-s)\gamma}{2} + \gamma^3 \,, \nonumber\\
\beta(j) &=& - j^2 + j\gamma^2  \,.
\label{oneloopbeta}
\end{eqnarray}
The phase diagram for the Bose-Fermi Kondo model thus shows a Kondo-screened
phase, a bosonic fluctuating phase, and a continuous quantum phase transition in between.
The RG flow is shown in Fig.~\ref{fig:bfkflow}.
Both intermediate-coupling fixed points are perturbatively accessible for small $(1\!-\!s)$;
it is likely that the structure of the phase diagram also applies to $s=0$ (i.e. $d=2$),
however, no accurate numerical calculations are available to date.
(As with the Bose Kondo model, for interacting bath bosons
the properties likely change smoothly for $-1<s<1$, whereas in the Gaussian
case a qualitative change is expected for $s\leq 0$.)
Both the critical and the bosonic fluctuating fixed points are
unstable w.r.t. breaking of SU(2) symmetry,
but the structure of the phase diagram is similar for both XY and Ising
symmetries, with the difference that in the Ising case the boson-dominated phase
corresponds to a strong-coupling fixed point~\cite{bfknew}.

If the bosonic bath is gapped, or its spectrum has exponent $s\geq 1$, then
the coupling to the bosons is (marginally) irrelevant, and the low-temperature
properties of (\ref{eq:bfk}) are that of the familiar fermionic Kondo model.
However, the Kondo temperature can be strongly renormalized due to the
competition from the bosonic bath.
An exception is the case of $s=1$ and Ising symmetry: here the bosonic coupling
drives a Kosterlitz-Thouless transition between a screened and unscreened
impurity \cite{hur}.

\subsection{Bose-Fermi Kondo model: large-$N$}

A multi-channel version of the Bose-Fermi Kondo model can be analyzed in a large-$N$ limit,
with a antisymmetric representation of SU($N$) for the impurity spin \cite{kirchner}.
The bosonic bath is treated as described for the Bose Kondo model (Sec.~\ref{sec:bklargen}),
and the fermionic bath is captured by the standard multi-channel NCA approach \cite{olivier}.
One obtains a set of coupled integral equations where the self-energy of the
auxiliary fermions now has contributions from both the bosonic and fermionic baths.
In the low-energy limit power-law solutions for the auxiliary-particle propagators
appear, corresponding to intermediate-coupling physics.

The phase diagram \cite{kirchner} is similar to the one obtained from the perturbative RG
described above, with the difference that large Kondo coupling produces a stable
intermediate-coupling fixed point corresponding to overscreening
(in contrast to the exactly screened situation of the single-channel model).
Importantly, $\omega/T$ scaling (\ref{scalchi1}) is obtained for all values of the bath
exponent $-1<s<1$.

\subsection{Results for observables}

The properties of the stable phases of the Bose-Fermi Kondo model
have already been discussed, see Sec.~\ref{sec:bkobs} for the boson-dominated
B-FL fixed point.
Notably, magnetic properties of the quantum critical point of (\ref{eq:bfk})
are very to the B-FL fixed point \cite{bfknew}:
the local susceptibility follows the scaling law (\ref{scalchi1}) with
an anomalous exponent $\eta_\chi > 0$ for $s<1$, and the
impurity susceptibility features a fractional Curie moment.
In addition, the conduction electron $T$ matrix at criticality,
discussed in Ref.~\onlinecite{MVMK}, is (only) logarithmically singular
for a metallic fermionic bath.
We note, however, that reliable numerical results for low temperatures
are lacking.

\subsection{Applications}

In the context of strongly correlated electron systems,
which often feature Fermi-liquid quasiparticles and strong spin fluctuations
at the same time, the question of the interplay between fermionic
and bosonic Kondo physics, as described by the Bose-Fermi Kondo model,
arises.
It has been proposed \cite{MVMK} that this interplay plays a role
in high-$T_c$ cuprates, where NMR experiments indicate that dilute impurity
moments are screened at optimal doping, but $T_K$ is essentially suppressed to
zero in underdoped compounds \cite{bobroff}.
A Fermi-Bose Kondo model, taking into account both the pseudogap density of states
of the Bogoliubov quasiparticles and the strong antiferromagnetic fluctuations,
provides a natural explanation: spin fluctuations increase with underdoping, thus
strongly reducing $T_K$ due to the vicinity to the boundary transition which
exists even in the absence of a bosonic bath \cite{MVMK}.

We note that other approaches to Kondo physics in correlated environments
have been put forward, which do not necessarily rely on the heuristic separation
of bosonic and fermionic bulk degrees of freedom.
Near a magnetic quantum critical point, ordered magnetization droplets form around
impurities and interact with the electronic environment \cite{millis}.
A related work on cluster Kondo physics \cite{shah} used an explicit microscopic
model for a magnetic nanostructure coupled to a metallic bath.
Further, it has been pointed out \cite{varma} that long-range critical bulk spin correlations can open
additional screening channels (even for an initially local impurity), resulting
in multi-channel Kondo physics at intermediate temperatures.
The class of cluster Kondo problems in disordered correlated environments
is closely related to the issue of quantum Griffiths behavior near
magnetic quantum critical points, with possible realizations in
chemically substituted heavy-fermion metals.

Various modifications of Bose-Fermi Kondo physics appeared recently in the
context of mesoscopic systems. Quantum dots coupled to ferromagnetic leads
involve both fermionic and bosonic baths plus an additional magnetic field
resulting arising from the lead polarization.
Charge fluctuations on a metallic island, described by Matveev's charge Kondo
effect \cite{matveev}, can be significantly modified by a noisy electromagnetic
environment, with the low-energy physics described by an anisotropic
Bose-Fermi Kondo model \cite{hur}.

Finally, we mention that the Bose-Fermi Kondo model has recently received
a lot of interest in the context of extended dynamical mean-field theory
\cite{edmft} where a lattice model is mapped onto a self-consistent impurity
model with both fermionic and bosonic baths.
Motivated by neutron scattering experiments \cite{schroder} on the
heavy-fermion compound $\rm CeCu_{6-x} Au_x$
which indicate momentum-independent critical dynamics at
an antiferromagnetic ordering transition,
a self-consistent version of the Bose-Fermi Kondo model has been
proposed to describe such ``local'' critical behavior.
In this scenario, the critical point of the lattice model is
mapped onto the FB-CR critical fixed point of the impurity model.


\section{Two-impurity Kondo models}
\label{sec:2imp}

Models of multiple impurities offer a new ingredient,
namely the exchange interaction between the
impurity spins, which competes with Kondo screening of the
individual impurities.
This inter-impurity interaction, which can lead to a magnetic
ordering transition in lattice models, arises both from direct
exchange and from the Ruderman-Kittel-Kasuya-Yosida (RKKY)
interaction mediated by the conduction electrons.

Let us consider a Kondo model with two impurity spins 1/2 ${\bf\SSS}_1,{\bf\SSS}_2$,
and a Hamiltonian written as
\begin{equation}
{\cal H} = \sum_i {\cal H}_{{\rm b},i}^{(\rm f)} + \sum_i \Jb {\bf\SSS}_i \cdot {\bf s}_i(0) +
{\cal H}_{12} \,.
\end{equation}
Each impurity spin is coupled to its individual bath ${\cal H}_{{\rm b},i}$,
i.e., there is a total of two screening channels,
and ${\cal H}_{12}$ contains the inter-impurity interaction.
Such models of two coupled impurities or two-level systems have been
investigated in a number of papers; the resulting phases and
phase transitions depend on the symmetry properties of
the bath and the couplings, and will be discussed in the following.

\subsection{Behavior for SU(2)-symmetric coupling}
\label{sec:2impsu2}

The case of SU(2)-symmetric direct exchange coupling between two
impurity $S=\frac{1}{2}$ spins,
\begin{equation}
{\cal H}_{12} = K\,{\bf\SSS}_1\cdot {\bf\SSS}_2 \,,
\end{equation}
has been investigated in detail in the literature \cite{2imp,2impnrg,2impcft}.
Two different regimes are possible as function
of the inter-impurity exchange~$K$:
for large antiferromagnetic $K$ the impurities combine to
a singlet, and the interaction with the conduction
band is weak, whereas for ferromagnetic $K$ the impurity
spins add up and are Kondo-screened by conduction electrons in the low-temperature limit.
It has been shown that these two parameter regimes are continuously
connected (without a $T=0$ phase transition) as $K$ is varied
in the generic situation without particle-hole symmetry.
Notably, in the particle-hole symmetric case one finds a transition
associated with an unstable non-Fermi liquid fixed
point~\cite{2impnrg,2impcft}.
This fixed point is somewhat similar to the one of the two-channel
Kondo model of Sec.~\ref{sec:2ck}, e.g., it features a residual entropy
of $\frac{1}{2} \ln 2$.

We note that different physics is encountered by coupling the
two-impurity system to a {\em single} conduction band channel only \cite{VBH}.
For ferromagnetic $K$ complete Kondo screening by the conduction electrons
is prohibited, resulting in an underscreened Kondo phase.
Upon variation of $K$ a Kosterlitz-Thouless-type transition between
a singlet and a doublet state occurs, associated with a second
exponentially small energy scale in the Kondo regime \cite{VBH}.

The physics becomes even richer if multi-channel physics
is combined with multi-impurity physics -- here, a variety
of fixed points including such with local non-Fermi liquid
behavior can be realized.

\subsection{Behavior for Ising coupling}
\label{sec:2impising}

The two-impurity model with Ising inter-impurity coupling,
\begin{equation}
{\cal H}_{12} = K_z \SSS_1^z \SSS_2^z \,,
\end{equation}
is qualitatively different from the SU(2)-symmetric one \cite{Andrei}.
It is particularly interesting, because
for large $|K_z|$ the two Ising-coupled spins form a magnetic
mini-domain which still contains an internal degree of freedom as the
ground state of $H_{12}$ is doubly degenerate
(in contrast to the inter-impurity singlet mentioned above).
For the case of antiferromagnetic $K_z$
the two low-energy states of the impurities (forming a pseudospin)
are $|\!\uparrow\downarrow\rangle$ and $|\!\downarrow\uparrow\rangle$.
The fate of this pseudospin degree of freedom
depends on the strength and asymmetry of the Kondo coupling $J$
between the spins and the bath electrons \cite{garst}.
On the one hand, for small Kondo couplings $J_\perp$, $J_z$ and large
$K_z$, tunneling between the two pseudospin configurations,
$|\!\uparrow\downarrow\rangle$ and $|\!\downarrow\uparrow\rangle$, is
supressed at low energies, i.e., the mini-domain is ``frozen'' as $T\to
0$, and the ground state entropy is $S_0=\ln 2$.
On the other hand, for small $K_z$ the two impurities are individually
Kondo screened,
resulting in a Fermi-liquid phase with vanishing residual entropy.
A quantum phase transition occurs for $K_z \sim \TK^{(1)}$,
where $\TK^{(1)}$ is the single-impurity Kondo temperature.
This phase transition is of Kosterlitz-Thouless type and
can also be tuned by varying the Ising component $J_z$ of the Kondo
coupling.
This leads to the presence of a second exponentially small
scale $T^\ast$, which is a collective Kondo temperature associated to
pseudospin screening.
The high-temperature $\ln 4$ impurity entropy is quenched in two stages:
first, at the scale $T^0 \approx K_z$,
the mini-domain ``forms'', quenching half of entropy; second
the strong fluctuations kill the remaining $\ln 2$ entropy at the much
lower scale $T^\ast$ \cite{garst}.

\subsection{Applications}

Experimentally, quantum dots provide an ideal laboratory
to study systems of two (or more) ``impurities'' -- note
that the local ``impurity'' states can arise either from
spin or charge degrees of freedom on each quantum dot.

In particular, a number of experiments have been performed
on multi-level or coupled quantum dot systems which can be
directly mapped onto models of two Kondo or Anderson
impurities \cite{wiel}.
Experiments reported in Ref.~\onlinecite{marcus} have used a setup of two
small quantum dots coupled to a larger metallic island
and various external leads.
Varying gate voltages allowed to tune the spin states of the
dots as well as the RKKY inter-dot coupling, and signatures of the
transition discussed in Sec.~\ref{sec:2impsu2} should be observable.

If charge degrees of freedom are employed for the Kondo effect \cite{matveev}
then two capacitively coupled dots present a promising realization
of the Ising coupling of Sec.~\ref{sec:2impising}.
A Kosterlitz-Thouless quantum phase transition will
occur upon variation of the capacitive inter-dot coupling.
Adding a small electron tunneling allows transport experiments
through this setup, and a universal conductance anomaly near the
transition has been predicted \cite{garst}.

Using recent advances in scanning probe techniques,
experimental realizations of multi-impurity models
using magnetic adatoms on metallic surfaces are possible \cite{cocumulti}.
Here, the spectral functions of the inidvidual impurities can be spectroscopically
measured, and the suppression of the Kondo effect for impurity dimers
has been demonstrated.
One interesting set of results is concerned with three Cr adatoms
forming a trimer -- this system has been proposed to show a stable fixed point
with local non-Fermi liquid behavior, provided that the threefold
spatial symmetry is unbroken \cite{3imp}.
Cluster of magnetic impurities, i.e., magnetic nanostructures, embedded in a
metallic environment represent a natural extension of these lines of thought.
An initial study of such a Kondo ``necklace'' model \cite{shah} showed
that multi-channel Kondo physics can be realized in a sizeable temperature
window.


\section{Multi-orbital Anderson models}
\label{sec:orb}

Similar to coupled impurities, multi-orbital or multi-level impurities
offer a larger set of local degrees of freedom which can undergo
a boundary phase transition.

Experimental realizations can be impurity atoms with
multiple low-lying crystal field states, offering bath charge (e.g. quadrupolar)
and spin degrees of freedom.
In the presence of orbital and spin degeneracy the resulting model
can show e.g. quadrupolar two-channel Kondo effect (see Sec.~\ref{sec:2ck}).
A distinct but related non-trivial fixed point has been found for
a two-orbital Anderson model with Hund's rule coupling \cite{leo}.

Alternatively, multi-level quantum dots can be tuned to study
Kondo physics beyond the simple Kondo model.
For instance, for two impurity levels and two electrons in the low-energy
sector the impurity can be close to a singlet--triplet
transition, which leads to a strongly enhanced Kondo temperature.
Other novel phenomena like a SU(4) Kondo effect have also been
discussed.
One can envision various quantum phase transitions in those devices
which have not yet been fully explored.


\section{Quantum--classical mapping}
\label{sec:qucl}

Significant progress in impurity physics has been made on the basis of
various mappings.
On the one hand, suitable transformations (e.g. bosonization, flow equations)
can lead to great simplifications.
On the other hand, Gaussian (non-interacting) bath degrees of freedom
can be formally integrated out, leading to a long-range self-interaction in
imaginary time direction for the impurity.
Discretizing imaginary time then leads to one-dimensional
statistical mechanics models with long-range interactions -- this
is a special case of the standard quantum--classical mapping for
quantum phase transitions.
(For bath particles with a relevant self-interaction the idea of the
quantum--classical mapping is inapplicable in general, as the bath cannot
be formally integrated out.)

\subsection{Ising symmetry}

Specifically, the spin-boson model (Sec.~\ref{sec:spb}) with
bath density of states $\omega^s$ yields an effective
impurity self-interaction
\begin{eqnarray}
{\cal S}_{\rm int} = \int d\tau d\tau' \sigma_z(\tau) g(\tau-\tau') \sigma_z(\tau')
\end{eqnarray}
with $g(\tau) \propto 1/\tau^{1+s}$ at long times.
The classical counterpart is the
one-dimensional Ising model \cite{leggett, weiss}
\begin{equation}
{\cal H}_{\rm cl} = - \sum_{\langle ij \rangle} J_{ij} S_i^z S_j^z + {\cal H}_{\rm SR}
\label{hcl}
\end{equation}
with interaction $J_{ij} = J/|i-j|^{1+s}$.
${\cal H}_{\rm SR}$ contains an additional generic short-range interaction
which arises from the transverse field, but is believed to be irrelevant
for the critical behavior \cite{fisher}.

For $s=1$ the quantum--classical mapping appears consistent:
both the ohmic spin-boson model and the $1/r^2$ Ising model show
a Kosterlitz-Thouless transition.
Notably, also the metallic Kondo model falls in this universality
class, which can be established using bosonization \cite{yuval,emery}.

The three models (classical Ising, anisotropic Kondo, spin-boson) display non-trivial
second-order transitions if the characteristic bath exponent is varied.
However, the transitions of the fermionic and bosonic quantum models
are no longer in the same universality class!
This has been shown by explicit calculations, see Sec.~\ref{sec:pgk}
for the pseudogap Kondo model and Sec.~\ref{sec:spb} for the sub-ohmic
spin-boson model.
For instance, the particle-hole pseudogap Kondo model shows a phase transition
only for $0<r<\frac 1 2$ (and has a complicated fixed point structure in the
asymmetric case, with a transition for all $r>0$),
whereas the spin-boson model has a phase transition for $0<s<1$.
These fundamental differences arise from the distinct character of the
bath degrees of freedom:
Linearly dispersing fermions in $(1+r)$ dimensions cannot be bosonized,
and a direct evaluation of the corresponding fermionic determinants is
not easily possible.
Contrary to metallic Kondo problems, the pseudogap Kondo model
is not conformally invariant, and conformal field theory techniques
cannot be applied.

The bosonic case appears simpler, and the formal mapping between the spin-boson model
and the classical Ising model (\ref{hcl}) is straightforward.
Indeed, the spin-boson model and the classical Ising model remain
equivalent in the range $\frac 1 2 < s <1$.
However, in Ref.~\onlinecite{quclfail} it has been shown that this equivalence breaks
down for $0<s<\frac 1 2$:
The classical model is effectively above its upper-critical dimension \cite{fisher},
given by $d=2s$, the transition takes mean-field exponents, and hyperscaling is
violated.
The spin-boson model, however, displays an interacting fixed point with non-trivial
exponents fulfilling hyperscaling, see Sec.~\ref{sec:spb}.
How can the quantum--classical mapping fail?
To establish the mapping a Trotter decomposition of the quantum partition
function is employed where the imaginary axis of length
$\beta = 1/T$ is divided into $N$ slices, leading to
an Ising chain (\ref{hcl}) with $N$ sites.
This procedure is exact when the limits $N\to\infty$ and $\beta\to\infty$
are taken in this order.
However, the limit $\beta/N\to 0$ leads to a {\em diverging} near-neighbor coupling
in the term ${\cal H}_{\rm SR}$ of the classical Ising model (\ref{hcl}) \cite{emery}.
This may in fact change the critical behavior of ${\cal H}_{\rm cl}$ (\ref{hcl}).
In other words, the quantum and classical problems are only equivalent
if the low-energy behavior of ${\cal H}_{\rm cl}$ is independent
of the order of limits \cite{emery}.
Apparently, these two limits cannot be interchanged for $s<1/2$,
and the naive quantum--classical mapping fails to describe the critical
properties of the strongly sub-ohmic spin-boson model.

\subsection{XY and SU(2) symmetries}

For bosonic quantum impurity models with continuous symmetry
our main knowledge arises from the RG calculations sketched in Sec.~\ref{sec:bk}.
For a quantum spin coupled to non-interacting vector bosons the naive
quantum--classical mapping would predict the equivalence to
a one-dimensional classical XY or Heisenberg model with long-range
interactions. These models are knows to display conventional
ordered and disordered phases.

In contrast, the RG analysis of Sec.~\ref{sec:bk} uncovered the existence
of a stable intermediate-coupling fixed point with non-trivial power laws;
the inclusion of an external field will drive a transition
to a disordered phase.
Thus, for XY and SU(2) symmetries even the stable (ordered) phase of the quantum model
is not correctly reproduced by the quantum--classical mapping for any value of $s<1$,
and the same likely applies to the transition properties.
A related observation appears in a SU($N$) version of the Bose-Fermi Kondo
model \cite{kirchner} where $\omega/T$ scaling is obeyed for all $s<1$,
contrary to the expectations from the classical model.

It is tempting to associate the general inapplicability of the mapping with the
presence of the Berry phase term describing the impurity spin dynamics --
this Berry phase is imaginary in the path integral representation and has no
classical analogue.
Thus we conjecture that the mapping works for a quantum {\em rotor} coupled to a
Gaussian bath, where the Berry phase term is absent.


\section{Summary}
\label{sec:summary}

We have reviewed a variety of zero-temperature critical points
in quantum impurity models, both with fermionic and bosonic baths.
Significant progress has been made in recent years, both in analytical and
numerical work, which has uncovered e.g. the existence of both lower-critical
and upper-critical dimensions and associated perturbative epsilon-expansions.
In addition, the validity of the naive quantum--classical mapping between
quantum impurity models and one-dimensional statistical mechanics models with
long-range interactions has been critically examined.

The impurity quantum phase transition have a variety of applications
in single-impurity models, e.g., for impurities embedded in correlated
electronic environments, and in quantum dot systems with potential applications
for spintronics.
In addition, effective impurity models appear in dynamical mean-field
approximations to correlated lattice models where they are supplemented
by self-consistency conditions \cite{dmft}.
Intermediate-coupling fixed points of impurity models have been proposed to describe
unconventional ``liquid'' phases of spin glass models \cite{sy} and
so-called local quantum critical points \cite{edmft}.

Interesting future directions include the study of non-equilibrium properties
near equilibrium quantum phase transitions, which are relevant, e.g., for
quantum dot transport experiments and for quantum decoherence processes.

\acknowledgments

The author is grateful to Z. Gulacsi for the smooth organization the 3rd Summer School
on Strongly Correlated Systems in Debrecen.
Most of the work reviewed here is based on fruitful collaborations with
R. Bulla, C. Buragohain, L. Fritz, M. Garst, W. Hofstetter, M. Kir\'{c}an,
Th. Prusch\-ke, A. Rosch, S. Sachdev, N. Tong, and M. Troyer.
The author also acknowledges illuminating discussions with
L. Balents, S. Florens, A.~C.~Hewson, D. Logan, N. Read, Q. Si,
D. Vollhardt, P. W\"olfle, and W. Zwerger.
Special thanks is to C. Lorenz for continued support during the writing
of this article.
This research was supported by the Deutsche Forschungsgemeinschaft
through the Center for Functional Nano\-structures Karls\-ruhe.


\end{document}